\documentstyle[12pt,epsf]{article}
\def\dfrac#1#2{{\displaystyle {#1 \over #2}}}

\def\dint{\displaystyle \int }

\newcommand{\ewxy}[2]{\setlength{\epsfxsize}{#2}\epsfbox[45 240 320 350]{#1}}

\newcommand{\ra}{\rightarrow }

\newcommand{\atc}{\alpha_{TC}}

\newcommand{\Ltc}{\Lambda_{TC}}
\newcommand{\Letc}{\Lambda_{ETC}}
\newcommand{\Lqcd}{\Lambda_{QCD}}
\newcommand{\LP}{\Lambda_{\Psi}}
\newcommand{\LQ}{\Lambda_Q}
\newcommand{\LL}{\Lambda_L}
\newcommand{\Li}{\Lambda_i}

\begin{document}

%%%%%%%%%%%%%%%%%%%%%%%%%%%%%%%%%%%%%%%%%%%%
% Title page
%%%%%%%%%%%%%%%%%%%%%%%%%%%%%%%%%%%%%%%%%%%%

\begin{titlepage}
\noindent BU HEP-95-16 \hfill July 1995\\
\noindent Napoli prep. DSF 21/95 \\

\begin{center}
\vspace{0.2cm}

{\LARGE\bf Production of neutral pseudo-Goldstone \\
bosons at LEP II and NLC in multiscale \\ \vspace{0.4cm}
walking technicolor models}

\vspace{0.8cm}

{\bf Vittorio~Lubicz$^a$ and Pietro~Santorelli$^b$} \\
\vspace{0.5cm}

{\it
$^a$Department of Physics, Boston University, \\
{\rm 590} Commonwealth Avenue, Boston, MA {\rm 02215}, USA \\
\vspace{0.2cm}
$^b$Dipartimento di Fisica, Universit\`a di Napoli ``Federico II'' and INFN,
\\ Sezione di Napoli, Mostra d'Oltremare, pad {\rm 19-20}, {\rm I-80125}
Napoli, Italy \\}
\vspace{0.5cm}

\end{center}

\abstract{ \par\noindent
Walking technicolor (WTC) models predict the existence of heavy neutral
pseudo-Gold\-stone bosons (PGBs), whose masses are typically expected to be
larger than 100 GeV. In this paper, we investigate the production and decay
of these particles at the high energy $e^+e^-$ experiments, LEP II and NLC.

We find that, in WTC models, the production of neutral PGBs can be
significantly enhanced, by one or two orders of magnitude, with respect to the
predictions of traditional (QCD-like) TC models. The origin of such an
enhancement is the existence of several low energy TC scales, that are likely
to appear in WTC theories. This could allow the PGBs to be observed even at
the energy and luminosity of the LEP II experiment. At LEP II, the PGBs are
expected to be produced in the $e^+e^- \ra P \gamma$ channel, and, possibly,
in the $e^+e^- \ra P e^+e^-$ channel, with a total rate that can be of the
order of several tenths per year. Due to the typical large values of PGB
masses, the relative branching ratios of PGB decays, in WTC theories, are
different from those predicted in traditional TC models. In particular, a
large fraction of these decays can occur in the $P \ra \gamma \gamma$ channel.
In considering the PGB production, at LEP II, we find that, in most of the
final states, the distinctive signatures of WTC events should allow the
Standard Model background to be reduced to a negligible level. We also find
that, at a 500 GeV NLC experiment, the production of neutral PGBs can occur
in several channels, and can be of the order of $10^3$ events per year.
Instead, when we consider traditional TC models, we find that no PGB are
typically predicted to be observed, both at LEP II and the NLC experiment.}
\vfill
\end{titlepage}

%%%%%%%%%%%%%%%%%%%%%%%%%%%%%%%%%%%%%%%%%%%%
% Sections
%%%%%%%%%%%%%%%%%%%%%%%%%%%%%%%%%%%%%%%%%%%%

\section{Introduction}

One of the most general predictions of any theory of dynamical electroweak
symmetry breaking and, in particular, of its most popular realizations,
technicolor (TC) \cite{Weinb}-\cite{FS} and extended technicolor (ETC) \cite
{DS,EL1} theories, is the existence of a large number of pseudo-Goldstone
bosons (PGBs). In all non-minimal TC models, for example, several tenths or
even hundreds of these particles are expected to exist, and the lightest of
these states should probably have sufficiently small masses to be produced
in current or presently planned experiments. Therefore, the experimental
searching of PGBs is a powerful tool in order to investigate the nature of
the electroweak symmetry breaking.

Over the last years, within the framework of traditional TC models, the
production of PGBs has been considered in a series of papers \cite{EHLQ}-%
\cite{VL}, both at hadron and lepton colliders. It is known, however, that,
in the attempt to correctly describe the ordinary fermion mass spectrum,
traditional TC models typically lead to large flavour changing neutral
currents (FCNC) \cite{EL1,DE}, which are not suppressed to a
phenomenologically acceptable level. For this reason, most of these models
are actually ruled out by the experimental data.

In addition, traditional (QCD-like) TC models present problems with
precision electroweak tests. The electroweak radiative correction parameter
$S$, for example, typically receives positive contributions in these
theories, and these contributions grow with the number of technifermion
doublets \cite{RCTC}. Experiments, however, seem to find $S$ to be very
small or even negative \cite{STU}.

Several years ago, walking technicolor (WTC) theories \cite{WTC} have been
proposed as a natural solution to the FCNC problem of TC. The WTC idea is
simple. In WTC theories, the running of the TC coupling constant, $\atc$,
between the TC scale $\Ltc$ and, approximately, the ETC scale $\Letc$, is very
slow. Consequently, the TC interactions remain strong enough over a large
range of momenta, and the values of the technifermion condensates,
renormalized at the scale $\Letc$, are significantly enhanced. This
enhancement also increases the values of the ordinary fermion and
technifermion masses, generated at the scale $\Letc$. Thus, in WTC theories,
$\Letc$ can be raised to several hundred TeV, and the ETC-generated FCNC are
properly suppressed.

In addition, WTC theories are not necessarily invalidated by precision
electroweak tests \cite{Lane}. In these theories, radiative corrections to the
Standard Model parameters are difficult to be reliably estimated \cite{Chiv},
mainly because the mass spectrum of WTC cannot be simply obtained by scaling
QCD data at higher energies. It has been argued, however, that in WTC the
value of the electroweak parameter $S$ could be smaller than the value
estimated in traditional TC models, and deviations from the Standard Model may
fall within current experimental bounds \cite{RCWTC}.

In this context, it is important to investigate whether signatures of WTC
theories might be detected in current or presently planned experiments. In
this paper, we have considered the production and decay of PGBs, in WTC
theories, at the high energy $e^+e^-$ colliders, LEP II and NLC. The
phenomenological analysis performed in this paper has been partially
inspired by the work of ref. \cite{LR}, in which WTC signatures, at hadron
colliders experiments, have been extensively studied. The relevant result of
ref. \cite{LR} is that, although most of the experimental signatures for TC
at hadron colliders have long been regarded as very difficult to detect \cite
{EHLQ}, however, the same consideration does not apply to models of WTC. In
this paper, we reach a similar conclusion: in WTC models, the PGBs production
at high energy lepton colliders can be significantly enhanced, by one or two
orders of magnitude, with respect to the predictions of traditional, QCD-like,
TC models. In particular, this enhancement could allow the PGBs to be observed
even at the energy and luminosity of the LEP II experiment.

In order to qualitatively understand this enhancement, we observe that two
features of WTC dynamics play, in this respect, a significant role. On one
hand, the enhancement of technifermion condensates, characteristic of WTC,
also raises the PGB masses, relative to their typical values in QCD-like TC
theories. Thus, one finds that, in WTC theories, the masses of the lightest
PGBs are expected to be larger than approximately 100 GeV, whereas typical
masses of the order of 40 GeV are predicted to be found in traditional TC
models \cite{EL1}. Incidentally, this also means that, if TC is walking, the
PGBs are likely to be too heavy to be produced at the LEP I or SLC
experiments.

On the other hand, several low energy TC scales are expected to exist in WTC
models. Multiscale TC models have been proposed in fact as a natural way to
implement a walking coupling \cite{EL2}. In multiscale TC models, the slow
running of $\alpha _{TC}$ is due to the existence of a large number of
technifermions. Typically, many technifermions belong to the fundamental
representation of the TC gauge group, while few technifermions may enter in
higher-dimensional representations. If this is the case, different values of
the technipion decay constants, $F_i$, are also expected to exist, depending
on the representations to which the corresponding technifermions $T_i$ belong
\cite{DRS}. Since the largest TC scale is bounded above by the characteristic
scale of electroweak symmetry breaking, $F_{max}\le 246$ GeV, then the
smallest scales can be relatively low, and the production of the corresponding
PGBs is consequently enhanced. This feature, i.e. the enhancement of the
cross section due to the existence of low TC scales, has been first pointed
out in ref. \cite{EL3}, in a study of top quark production at $p\bar p$
colliders.

In considering the production of PGBs in $e^+e^-$ collisions, we will
concentrate in this paper on those PGBs that are electrically neutral and
color singlet. There are in fact several reasons why these states are worthy
of interest. First of all, the neutral and colorless PGBs are expected to be
the lightest ones, since their masses do not receive contributions from the
standard electroweak and strong interactions. In addition, the neutral PGBs
can be singly produced in $e^+e^-$ collisions, therefore allowing the probe of
higher mass scales. Finally, the couplings of neutral PGBs to ordinary
fermions and gauge bosons can be expressed in a form that is, to some extent,
model independent \cite{VL}. This feature is particularly interesting, since
no completely realistic TC model has been constructed so far. All the model
dependence of the neutral PGB couplings to ordinary particles, fermions and
gauge bosons, is explicitly included in the values of the relevant energy
scales of the theory (the pseudoscalar decay constants of technipions), in the
dimension of the TC gauge group, and in few model dependent couplings, which,
however, are typically expected to be of order one. Thus, we are able to
explore quite general predictions of the theory for the various production and
decay rates.

Nevertheless, the precise values of cross sections, for PGB production, still
depend on the details of the specific model. For this reason, in discussing
our results, we have allowed a large range of variability for the PGB masses
and considered different values of couplings. In the lack of a completely
satisfactory TC model, we will not go, in this paper, in the details of the
several theoretical proposals. Our main intent is to show that the PGBs of WTC
theories are expected to be produced, in presently planned experiments, for a
reasonable range of model parameters, so that they are worth of an
experimental research.

For illustrative purposes, we have considered in this paper a simple model of
WTC, namely the multiscale TC model proposed by Lane and Ramana (LR) in ref.
\cite{LR}. The three neutral PGBs, entering in this model, are coupled with
different strengths to the ordinary particles. Therefore, the analysis of this
model will allow us to investigate different possible scenarios. In addition,
in order to compare the WTC predictions with those of traditional, single
scale, TC models, we will also consider in this paper the popular, one family,
TC model proposed by Farhi and Susskind (FS) in ref. \cite{FSM}.

Our results indicate that, at the LEP II experiment, the main contribution to
the cross section of neutral PGB production is expected to come from the $e^+
e^- \ra P\gamma$ channel. In this channel, by taking into account the
experimental cuts and reconstruction efficiencies, we find that several tenths
of PGBs can be produced per year, assuming an integrated luminosity of 500
pb$^{-1}$. Possibly, a smaller PGB production can be also observed in the $e^+
e^- \ra P e^+e^-$ channel. As far as the decay modes of these particles are
concerned, we find that the predictions of WTC theories can be significantly
different from those obtained in traditional TC models. In particular, due to
the typical large values of PGB masses, a large fraction of PGB decays can
also occur in the $P \ra \gamma \gamma$ channel, besides the
``traditional" decays into a gluon or a bottom quark pair. We then show that,
at LEP II, the experimental signature of WTC events is, in most of these
channels, quite distinctive, thus allowing the Standard Model background to be
reduced to a negligible level.

At a 500 GeV NLC experiment, the production of neutral PGBs, predicted in WTC
models, is significantly larger, and it is estimated to be of the order of
$10^3$ events per year, by assuming an integrated luminosity of $10^4$
pb$^{-1}$. This production is expected to occur mainly into the $e^+e^-\ra
P \gamma$, $e^+e^- \ra P Z^0$ and $e^+e^- \ra P e^+e^-$ channels. Instead, we
find that, within the framework of traditional TC models, no PGBs are
typically predicted to be observed, both at LEP II and the NLC experiment.

In the rest of this paper we will discuss our results in more detail. The LR
model of WTC will be briefly reviewed in section 2, and the masses of the
lightest neutral PGBs in this model will be also estimated. In section 3 we
will discuss the general form of the neutral PGB couplings to ordinary
particles. The values of the model dependent constants, entering in these
couplings, will be computed in the particular case of the LR model. In section
4, we will consider the expected decay modes of PGBs and give the results for
the various production cross sections. Finally, in section 5, we will present
our conclusions.

\section{The LR model of WTC}

So far, a completely realistic and self-consistent TC model has not yet been
constructed. In particular, the LR model we are going to discuss presents ETC
gauge anomalies and predicts a too much large number of ordinary quarks and
leptons. However, the model reproduces the major aspects of a typical,
``quasi-realistic" model of WTC, by including several species of
technifermions, several scales of technifermion chiral symmetry breaking,
and $SU(2)$ isospin breaking, that accounts for the up-down mass splittings
of ordinary fermions. In addition, the model is simple enough, and allows
for a detailed estimates of technifermion and technipion masses. All these
features turn out to be important for our purposes. In this section, we will
review those aspects of the LR model that are relevant for this paper, and we
refer the interested reader to ref. \cite{LR} for more details. In addition,
we will compute the values of neutral PGBs masses in this model, to be used in
our subsequent phenomenological analysis.

The LR model is based on the ETC gauge group:

\begin{equation}
\label{ETCLR} SU(N_{ETC})_1\otimes SU(N_{ETC})_2
\end{equation}

\noindent
where $N_{ETC}=N_{TC}+N_C+N_L$. $N_{TC}$ represents the number of
technicolors, $N_C$ the number of ordinary colors and $N_L$ the number of
fermion flavours. The number of colors is fixed to the physical value,
$N_C=3$, while the number of technicolors, $N_{TC}$, and the number of
flavours, $N_L$, are chosen to be the minimal ones to guarantee the walking of
TC coupling constant. In ref. \cite{LR}, they find that this condition
corresponds to $N_{TC}=N_L=6$, and we will assume these values throughout
this paper.

The dynamical symmetry breaking of the ETC group proceeds through two
different steps. A first breaking occurs at the scale $M_A$, when the group
$SU(N_{ETC}) _1 \otimes SU(N_{ETC})_2$ is broken down to the diagonal subgroup
$SU(N_{ETC}) _{1+2}$. At a lower scale, $M_V$, a further breaking occurs, and
the residual gauge symmetry becomes ${\cal G}=SU(N_{TC}) \otimes SU(3)_C
\otimes SU(N_L)$. The scale $M_V$ and $M_A$ are estimated to be of the order
of $M_V \simeq 100$ TeV and $M_A \simeq 400$ TeV.

All technifermions and ordinary fermions in the model can be classified
according to the corresponding representations of the gauge group ${\cal G}$
to which they belong. In particular, the model contains three different
species of technifermions. One doublet of color-singlet technifermions,

\begin{equation}
\label{Psi} \Psi =(\Psi _U,\Psi _D)
\end{equation}

\noindent
belonging to the antisymmetric second-rank tensorial representation of the TC
gauge group; one doublet of color-triplet techniquarks,

\begin{equation}
\label{Q} Q_c=(U_c,D_c)
\end{equation}

\noindent
with $c=1,2,3$; and $N_L$ doublets of color-singlet technileptons,

\begin{equation}
\label{L} L_l=(N_l,E_l)
\end{equation}

\noindent
with $l=1,\ldots ,N_L$. Both $Q$ and $L$ transform as the fundamental
representation of TC gauge group $SU(N_{TC})$.

As far as the ordinary fermions are concerned, their number in the LR model
is unrealistically large. With $N_L=6$, the model contains 6 doublets of
quarks, one doublets of antiquarks and $N_L\,(N_L-1) \,/\,2=15$ doublets of
ordinary leptons. Clearly, in order to consider the model predictions for
physical processes with ordinary fermions entering in the initial and final
states, we will be forced to introduce, in this respect, some approximations.
However, we postpone this discussion to the last section.

The LR model of WTC is an example of multiscale TC model. Since the
technifermions $\Psi$ and $Q,L$ belong to inequivalent representations of
the TC group, they are associated to different scales $(\Li)$ of TC chiral
symmetry breaking \cite{DRS}. Consequently, three different values of
technifermion decay constants $(F_i)$ are also expected to occur. They are
constrained by the condition:

\begin{equation}
\label{FPLR} F_{\pi} \equiv \sqrt{F_\Psi ^2\,+\,3\,F_Q^2\,+\,N_L\,F_L^2}=246\
{\rm GeV}
\end{equation}

\noindent
that guarantees the correct physical values are assigned to the $W^{\pm}$ and
$Z^0$ boson masses.

In ref. \cite{LR}, the values of $F_i$ and $\Li$ are computed by scaling the
ratio $\Li/ F_i$ from the QCD ratio $\Lqcd /f_{\pi}$. Two possible scaling
rules have been considered, that differently take into account the dependence
on the dimensionalities $d_i$ of the $SU(N_{TC})$ technifermion
representations. The first rule $(A)$ assumes that $F_i$ scales like $\Li
\sqrt{d_i}$ (according to an expansion in $1/d_i$), while the second rule
$(B)$ assumes that $F_i\,/\,\Li$ is independent on $d_i$. In this way, two
different sets of model parameters are obtained. The corresponding values of
TC scales and decay constants are shown in table \ref {tab:C}.
%_________________________________________________________________________
\begin{table} \centering
\renewcommand{\arraystretch}{1.5}
\begin{tabular}{|c||c|c|c||c|c|c|} \hline
& $\LP$ & $\LQ$ & $\LL$ &
  $F_\Psi$ & $F_Q$ & $F_L$ \\ \hline \hline
$A$ & $ 428 $ & $ 83  $ & $ 82  $ & $ 231 $ & $ 29 $ & $ 28 $ \\
$B$ & $ 876 $ & $ 177 $ & $ 172 $ & $ 212 $ & $ 43 $ & $ 41 $ \\ \hline
\end{tabular}
\renewcommand{\arraystretch}{1.0}
\caption{\it Values (in GeV) of the TC scales, $\Li$, and decay constants,
$F_i$, in the LR model, for the two sets, $A$ and $B$, of model parameters.
\label{tab:C}}
\end{table}
%_________________________________________________________________________
Note that a small splitting between the two lowest scales, $\LQ$ and $\LL$,
occurs, due to the weak effects of QCD color interactions at the TC scale.

In both sets of parameters, $A$ and $B$ of table \ref{tab:C}, the decay
constant $F_{\Psi}$ turns out to be quite close to the electroweak symmetry
breaking scale, $F_{\pi}=246$ GeV. Eq. (\ref{FPLR}) then implies that the
other two scales, in the model, must be very low, of the order of few tenths
of GeV. The existence of such low scales is expected to be a general feature
of multiscale WTC models.

When we neglect the ETC interactions, we find that technifermions of the LR
model have a large chiral flavour symmetry group:

\begin{equation}
\label{ChiralLR}
\left[ SU(2)_L\otimes SU(2)_R\right]_\Psi \otimes
\left[ SU(2N_L+6)_L\otimes SU(2N_L+6)_R\right]_{Q,L}
\end{equation}

\noindent
By effect of TC interactions, these chiral symmetries are spontaneously broken
to the corresponding vector subgroup, and the symmetry breakdown produces
$3+(2N_L+6)^2-1=326$ Goldstone bosons. Three of these particles become the
longitudinal components of the $W^{\pm }$ and $Z^0$ bosons. The remaining 323
states represent true PGBs, that acquire mass mainly from the ETC interactions.

The large scale hierarchy in the model (see table \ref{tab:C}) implies that
the three would-be Goldstone bosons are mainly constituted by the
technifermions $\Psi$. Precisely, by denoting with $\alpha^2 _\Psi$, $\alpha^2
_Q$ and $\alpha^2 _L$ the $\Psi-$, $Q-$ and $L-$content of the absorbed
Goldstone bosons, we have:

\begin{equation}
\label{bi}
\alpha _\Psi = \frac{F_\Psi }{F_\pi} \, , \quad
\alpha _Q    = \frac{\sqrt{3}\,F_Q}{F_\pi} \, , \quad
\alpha_L     = \frac{\sqrt{N_L} \,F_L}{F_\pi}
\end{equation}

\noindent
with $F_\pi=246$ GeV (eq. (\ref{FPLR})). Using the values of decay constants
given in table \ref{tab:C}, we then find that the techniquark and technilepton
content of the absorbed technipions is approximately of the order of $\alpha^2
_Q + \alpha^2 _L \simeq 10-25 \%$.

In the following, in order to simplify our analysis, we will neglect this
mixing, and we will assume that the three absorbed technipions are only
constituted by the technifermions $\Psi$, i.e. $\alpha_{\Psi} \simeq 1$. In
this limit, the true PGBs, observed in the spectrum, are those particles only
constituted by the $Q$ and $L$ technifermions, i.e. the states generated by
the dynamical breakdown of the $SU(2N_L+6)_L\otimes SU(2N_L+6)_R$ chiral
symmetry.

Three of the physical PGBs, in the model, are electrically neutral and belong
to a singlet of the $SU(3)$ color group and to a singlet of the $SU(N_L)$
flavour group. The corresponding fields are proportional to the following
linear combinations:

\begin{equation}
\label{P30LR}
\renewcommand{\arraystretch}{1.3}
\begin{array}{l}
P_Q^3\sim \overline{U}_c\,\gamma ^5\,U_c\,-\,\overline{D}_c\,\gamma ^5\,D_c \\
P_L^3\sim \overline{N}_l\,\gamma^5\,N_l\,-\,\overline{E}_l \,\gamma ^5\,E_l \\
P^0\sim N_L\,(\overline{U}_c\,\gamma ^5\,U_c\,+\, \overline{D}_c\,\gamma^5 \,
D_c)\, -\,N_C\,(\overline{N}_l\,\gamma^5\,N_l \,+\,\overline{E}_l\, \gamma^5\,
E_l)
\end{array}
\renewcommand{\arraystretch}{1.0}
\end{equation}

\noindent
where repeated color ($c=1,2,3$) and flavour ($l=1,\ldots ,N_L$) indexes are
summed over. These PGBs are the particles we are interested in. By carrying
neither color nor electric charge, they represent the lightest states of the
PGB mass spectrum, and they can be produced with a larger probability in the
$e^+e^-$ collision experiments.

In order to consider these particles in our phenomenological analysis, we now
estimate the size of their masses. The neutral states of eq. (\ref{P30LR}) can
only receive mass from the ETC interactions, which explicitly break
technifermion chiral symmetries. We can parameterize the ETC symmetry breaking
interactions in terms of an effective Hamiltonian, containing the ETC
generated $Q$ and $L$ technifermion masses:

\begin{equation}
\label{HETCLR} H_{ETC}=m_U\,\overline{U}U\,+\,m_D\,\overline{D}D\,+\,m_N\,
\overline{N} N\,+\,m_E\,\overline{E}E
\end{equation}

\noindent
Then, at the first order in the symmetry breaking Hamiltonian, the PGB mass
matrix is given by the Dashen formula \cite{Dash}:

\begin{equation}
\label{DashF} \left(M^2_P\right)_{ab}=\dfrac 1{F_P^2}\,\langle 0\mid
\left[Q_5^a,\left[ Q_5^b, H_{ETC} (0)\right] \,\right] \mid 0\rangle
\end{equation}

\noindent
where $F_P$ is the decay constant ($F_P=F_Q\simeq F_L$), $H_{ETC}$ is the
symmetry breaking Hamiltonian (eq. (\ref{HETCLR})) and $Q_5^a$ are the axial
charges associated with the flavour chiral symmetry group.

The explicit evaluation of the commutators in the Dashen formula is cumbersome
but straightforward. One then finds that the mass matrix of the three neutral
PGBs is not diagonal, since the mass splittings, $(m_U-m_D)$ and $(m_N-m_E)$,
give raise to a mixing among the states of eq. (\ref{P30LR}). In a first
approximation, we can neglect this mixing, by simply neglecting the
off-diagonal elements of the PGB mass matrix. We then find that the masses of
the three light neutral PGBs in the model are given by the expressions:

\begin{equation}
\label{MP03LR}
\renewcommand{\arraystretch}{2.0}
\begin{array}{c}
M_{3Q}^2\,F_Q^2=
(m_U+m_D)_{\LQ}\,\langle \, \overline{Q}Q\, \rangle _{\LQ}\ , \qquad
M_{3L}^2\,F_Q^2=
(m_N+m_E)_{\LL}\,\langle \,\overline{L}L\,\rangle _{\LL}\ , \\
M_0^2\,F_Q^2=
\left( \dfrac{N_L}{N_L+N_C}\right) (m_U+m_D)_{\LQ} \, \langle \,
\overline{Q}Q\,\rangle _{\LQ}\,+\ \left( \dfrac{N_C}{N_L+N_C} \right)
(m_N+m_E)_{\LL}\,\langle \,\overline{L}L\,\rangle _{\LL}
\end{array}
\renewcommand{\arraystretch}{1.0}
\end{equation}

\noindent
where $\langle \,\overline{Q}Q\,\rangle$ and $\langle \,\overline{L} L\,
\rangle $ are the techniquark and technilepton condensates. The values of
these condensates and the values of the technifermion masses have been
evaluated in ref. \cite{LR}, and they are presented in table \ref{tab:A} for
the two sets, $A$ and $B$, of model parameters.
%_________________________________________________________________________
\begin{table} \centering
\renewcommand{\arraystretch}{1.5}
\begin{tabular}{|c||c|c||c|c|c|c|} \hline
& $\langle \, \overline{Q}Q\,\rangle$ &
  $\langle \, \overline{L}L\,\rangle$ &
  $m_U$ & $m_D$ & $m_N$ & $m_E$
\\ \hline \hline
$A$ & $(69)^3 $ & $(66)^3 $ & $136$ & $22$ & $61$ & $13$ \\
$B$ & $(114)^3$ & $(109)^3$ & $ 92$ & $14$ & $43$ & $9 $ \\ \hline
\end{tabular}
\renewcommand{\arraystretch}{1.0}
\caption{\it Values of the technifermion condensates (in unit of (GeV)$^3$)
and masses (in GeV), evaluated at the TC scale $\LQ\simeq \LL$, in the LR
model, for the two sets, $A$ and $B$, of model parameters.
\label{tab:A}}
\end{table}
%_________________________________________________________________________
By substituting these values in eq. (\ref{MP03LR}), we then obtain (in GeV):

\begin{equation}
\label{MP03N}
\renewcommand{\arraystretch}{1.3}
\begin{array}{llll}
(A)\quad & M_{3Q}=247\,,\quad & M_{3L}=166\,,\quad & M_0=223 \\
(B)\quad & M_{3Q}=293\,,\quad & M_{3L}=200\,,\quad & M_0=266
\end{array}
\renewcommand{\arraystretch}{1.0}
\end{equation}

This calculation can be now repeated by correctly taking into account the
mixing among the three PGBs. In this case, by denoting with $M_{P1},\ M_{P2}$
and $M_{P3}$ the eigenvalues of the PGB mass matrix, we find (in GeV):

\begin{equation}
\label{MPAVN}
\renewcommand{\arraystretch}{1.3}
\begin{array}{llll}
(A)\quad & M_{P1}=304\,,\quad & M_{P2}=118\,,\quad & M_{P3}=173 \\
(B)\quad & M_{P1}=362\,,\quad & M_{P2}=137\,,\quad & M_{P3}=205
\end{array}
\renewcommand{\arraystretch}{1.0}
\end{equation}

\noindent
Thus, the masses of the three neutral PGBs, in the LR model, are in the range
between 100 and 350 GeV. The main source of uncertainty in this estimate comes
from the scaling from QCD of the PGB decay constants and technifermion
condensates. For this reason, the results of eqs. (\ref{MP03N}) and
(\ref{MPAVN}) must be considered as purely indicative. However, it is also
reasonable to assume that they correctly represent the typical order of
magnitude of the lightest PGB masses in the framework of a general multiscale
WTC models.

\section{The couplings of neutral PGBs to ordinary particles}

In TC/ETC theories, the couplings of the neutral PGBs to ordinary fermions and
gauge bosons are, to some extent, model independent. All the model dependence
of these couplings is included in the values of the relevant energy scales of
the theory, the pseudoscalar decay constants of technipions, in the dimension
of the TC gauge group, $N_{TC}$, and in two classes of constants which,
however, are typically expected to be of order one. In this section, we will
write down the general form of these couplings and we will compute the values
of the model dependent constants in the particular case of the LR model.

\subsection{Couplings of neutral PGBs to gauge bosons}

Let us first define the couplings of neutral PGBs with two arbitrary gauge
bosons of the Standard Model.

At the energy scales smaller than the typical TC scale, $\Ltc$, the couplings
between a PGB $P$ and two gauge bosons, $B_1$ and $B_2$, are controlled by the
ABJ anomaly \cite{ABJ}. By adopting a convenient parameterization, these
couplings can be written in the form \cite{VL}:

\begin{equation}
\label{APBB} \dfrac{1}{(1+\delta _{B_1B_2})}\,\left( \dfrac{\alpha\, d_{TC}\,
A_{PB_1B_2}}{\pi F_P\sqrt{n/2}}\right) \,P \,\epsilon _{\lambda \mu \nu \rho}
\, (\partial ^\lambda B_1^\mu )\,(\partial ^\nu B_2^\rho )
\end{equation}

\noindent
where $P$, $B_1^\mu $ and $B_2^\mu $ represent the field operators of the PGB
and the two gauge bosons $B_1$ and $B_2$ respectively. $F_P$ is the PGB decay
constant, $n$ is the dimension of the chiral flavour symmetry group and $d
_{TC}$ is the dimensionality of the TC representation to which the
technifermions $T$, constituting the PGB $P$, belong. In the LR model, for
instance, when we consider the techniquark and technilepton sector, we have
$n=2(N_L+3)$ and $d_{TC}=N_{TC}$. Since technifermions entering in the
fundamental representation of the group $SU(N_{TC})$ is a very usual
condition in TC models, in the following we will always consider the case
$d_{TC}=N_{TC}$. We also define $\alpha$, in eq. (\ref{APBB}), to be $e^2/4
\pi $ if $B_1$ and $B_2$ are electroweak gauge bosons, and equal to the
strong coupling constant, $\alpha _s$, if $B_1$ and $B_2$ are QCD gluons.

The coupling $A_{PB_1B_2}$, in eq. (\ref{APBB}), is a group theoretical
factor. It is defined by the relation:

\begin{equation}
\label{Esse}
4\pi \alpha\,\sqrt{2/n}\,A_{PB_1B_2} = g_1 g_2 \,{\rm Tr} \left[ Q_P\,
(\{ Q_V^1 , Q_V^2 \}+ \{ Q_A^1 , Q_A^2\})\,\right]
\end{equation}

\noindent
where $g_i$ is the gauge coupling of the boson $B_i$ and $Q_{V,A}^i$ are the
corresponding vector and axial charges. $Q_P$ is the axial charge of the PGB
$P$. The factor $\sqrt{2/n}$, entering in eq. (\ref{Esse}), approximately
takes into account the dependence of the anomalous couplings on the dimension
of the flavour group, in such a way that the coupling $A_{PB_1B_2}$ is
expected to be a constant of order one, independently on the particular TC
model one is considering \cite{VL}.

It is useful to verify such approximate model independence by considering some
particular cases. In the $\pi^0 \ra 2\gamma$ decay of QCD, we have $d_{TC}=3$
(in this case, this is the number of ordinary colors), $n=2$ (the dimension
of the flavour symmetry group) and $A_{\pi^0 \gamma\gamma}=1/3$.

In the traditional, one family, TC model, introduced by Farhi and Susskind in
ref. \cite{FSM}, there exist one doublet of color-triplet techniquark, $Q_c=
(U_c,D_c)$, and one doublet of color-singlet technileptons, $L=(N,E)$. In this
model, the spontaneous breakdown of the $SU(8)_L \otimes SU(8)_R$ chiral
symmetry of technifermions results in 63 Goldstone bosons, two of which, $P^0$
and $P^3$, are neutral and colorless:

\begin{equation}
\label{P30FS}
\renewcommand{\arraystretch}{1.3}
\begin{array}{l}
P^0\sim (\overline{U}_c\,\gamma ^5\,U_c\,+\, \overline{D}_c\,\gamma^5 \, D_c)
\, -\, 3\,(\overline{N}\,\gamma^5\,N \,+\,\overline{E}\, \gamma^5\, E) \\
P^3\sim (\overline{U}_c\,\gamma ^5\,U_c\,-\,\overline{D}_c\,\gamma ^5\,D_c)
\, - 3\,(\overline{N}\,\gamma^5\,N\,-\,\overline{E} \,\gamma ^5\,E)
\end{array}
\renewcommand{\arraystretch}{1.0}
\end{equation}

\noindent
They belong to an isospin singlet and triplet respectively, and they are
expected to be the lightest PGB states contained in the model. The values of
the corresponding model dependent constants, $A_{PB_1B_2}$, for $B_1B_2$ equal
to $ZZ$, $Z\gamma$, $\gamma \gamma$ and a gluon pair, $g^ag^b$, are shown in
the upper side of table \ref{tab:B}. In this table, together with the
analytical expressions, we also give the corresponding numerical values,
obtained by using $s_W^2=0.23$, where $s_W$ is the sine of the Weinberg angle.
{}From table \ref{tab:B}, we see that all these constants are approximately of
order one.
%_________________________________________________________________________
\begin{table} \centering
\renewcommand{\arraystretch}{2.0}
\begin{tabular}{|c||c|c||c|c||c|c||c|c||} \hline
{\bf FS}
 & \multicolumn{2}{c||}{$A_{PZZ}$}
 & \multicolumn{2}{c||}{$A_{PZ\gamma}$}
 & \multicolumn{2}{c||}{$A_{P\gamma \gamma}$}
 & \multicolumn{2}{c||}{$A_{P g^a g^a}$}
 \\ \hline \hline
$ P^0 $
&$ -\dfrac{4}{3\sqrt{3}}\,t_W^2 $ & $-0.23$
&$  \dfrac{4}{3\sqrt{3}}\,t_W $   & $ 0.42$
&$ -\dfrac{4}{3\sqrt{3}} $        & $-0.77$
&$  \dfrac{1}{\sqrt{3}} $         & $ 0.58$
\\
$ P^3 $
&$ -\dfrac{4}{\sqrt{3}}\,\dfrac{(1-2s_W^2)}{2\,c_W^2} $    & $-0.81$
&$  \dfrac{4}{\sqrt{3}}\,\dfrac{(1-4s_W^2)}{4\,s_W\,c_W} $ & $ 0.11$
&$  \dfrac{4}{\sqrt{3}} $                                  & $ 2.31$
&$  0 $                                                    & $ 0.00$
\\ \hline \hline
{\bf LR}
 & \multicolumn{2}{c||}{$A_{PZZ}$}
 & \multicolumn{2}{c||}{$A_{PZ\gamma}$}
 & \multicolumn{2}{c||}{$A_{P\gamma \gamma}$}
 & \multicolumn{2}{c||}{$A_{P g^a g^a}$}
 \\ \hline \hline
$ P^0 $
&$ -\dfrac{4\sqrt{2}}{3}\,t_W^2 $ & $-0.56$
&$  \dfrac{4\sqrt{2}}{3}\,t_W $   & $ 1.03$
&$ -\dfrac{4\sqrt{2}}{3} $        & $-1.89$
&$  \sqrt{2}$                     & $ 1.41$
\\
$ P_Q^3 $
&$ -\sqrt{3}\,\dfrac{(1-2s_W^2)}{2\,c_W^2} $     & $-0.61$
&$  \sqrt{3}\,\dfrac{(1-4s_W^2)}{4\,s_W\,c_W} $  & $ 0.08$
&$  \sqrt{3} $                                   & $ 1.73$
&$  0 $                                          & $ 0.00$
\\
$ P_L^3 $
&$  3\sqrt{6}\,\dfrac{(1-2s_W^2)}{2\,c_W^2} $    & $ 2.58$
&$ -3\sqrt{6}\,\dfrac{(1-4s_W^2)}{4\,s_W\,c_W} $ & $-0.35$
&$ -3\sqrt{6} $                                  & $-7.35$
&$  0 $                                          & $ 0.00$
\\ \hline
\end{tabular}
\renewcommand{\arraystretch}{1.0}
\caption{\it Expressions and numerical values of the constants $A_{PB_1B_2}$
in the FS and LR models. The symbols $s_W$, $c_W$ and $t_W$ represent the
sine, cosine and tangent of the Weinberg angle respectively, and the numerical
values have been obtained by using $s_W^2=0.23$.
\label{tab:B}}
\end{table}
%_________________________________________________________________________

Clearly, some particular deviations from unity can always occur. For example,
the coupling $A_{P^3 Z\gamma}$, of the PGB $P^3$, turns out to be quite small,
being proportional to the small combination $(1-4s_W^2)$. As observed in refs.
\cite{Ran1,Ran2}, for all the PGBs belonging to a triplet representation of
the isospin group, this proportionality is a general consequence of the
underlying custodial $SU(2)$ symmetry. The same symmetry also produces the
vanishing of the $P^0$ and $P^3$ couplings to a $W^+W^-$ boson pair and the
vanishing of the coupling between the isovector $P^3$ and a QCD gluon pair.

In the lower side of table \ref{tab:B}, we give the values of the couplings
$A_{PB_1B_2}$ for the three light neutral PGBs of the LR model. Note how the
structure of these coefficients is fixed by the custodial $SU(2)$ symmetry.
The couplings of the PGBs $P^3_Q$ and $P^3_L$, of the LR model, are
proportional to those of the PGB $P^3$ in the FS model, and similarly for the
couplings of the two isoscalar $P^0$'s. Moreover, we find that, also in LR
model, all these couplings are of the order of one, thus supporting the
statement of their approximate independence on the particular choice of the TC
model.

\subsection{High energy corrections to anomalous couplings}

The form of the anomalous coupling, considered in eq. (\ref{APBB}), is
expected to be exactly valid only in the limit in which the energy scales and
the masses of the particles, involved in the process, are smaller than the
typical TC scale, $\Ltc$. In WTC theories, however, the masses of the PGBs are
typically of the order of $\Ltc$. Moreover, we are going to consider in this
paper physical processes occurring at the energy scales of the LEP II and NLC
experiments, namely 200 and 500 GeV respectively. Even these scales are both
of the order, or even larger, than $\Ltc$. Thus, in computing the
corresponding cross sections, the use of the anomalous couplings of eq.
(\ref{APBB}) might be questionable.

The fact that the couplings of eq. (\ref{APBB}) loose their validity in the
high energy limit is also indicated by the following observation. Let us
consider any process in which a virtual gauge boson, $B_1$, is produced in the
s-channel, and it is coupled, via the anomaly, to a PGB and a second gauge
boson, $B_2$. An example of this process is $e^+e^- \ra \gamma^* \ra P\gamma$.
Since the couplings of eq. (\ref{APBB}) turn out to be inversely proportional
to the pseudoscalar decay constant $F_P$, one can argue, from simple
dimensional arguments, that, in the high energy limit, the corresponding cross
sections must behave like a constant. Indeed, this is just what we will find,
after an explicit calculation, in section 4 (see eq. (\ref{SeePB}) below). The
same problem arises in QCD, if one considers, in the high energy limit, the
$\pi^0$ production in the $e^+e^- \ra \pi^0 \gamma$ channel. Thus, at high
energies, the anomalous couplings of eq. (\ref{APBB}) violate the unitarity of
the theory.

By considering parity and Lorentz invariance, it is easy to realize that the
general form of the couplings between a PGB and two vector gauge bosons is
still given by eq. (\ref{APBB}) times an invariant form factor.

In QCD, or in a strongly interacting TC theory, this ``anomalous" form factor
can be only evaluated with a full non-perturbative calculation of the two
triangle diagrams, that, in the low energy limit, originate the anomaly. In
these diagrams, the quark or technifermion lines, coupled in a vertex to a
pseudoscalar source, give rise, by effect of their mutual strong interactions,
to the propagation of the PGB field.

In the lack of a non-perturbative calculation, we try to estimate the high
energy behaviour of the triangle diagrams in the context of a phenomenological
model. We have considered a simple version of the linear $\sigma$ model,
containing one charged fermion field (the proton), one pseudoscalar field (the
neutral pion) and the scalar $\sigma$ field. In this model, the effective
coupling between the pion and two vector gauge boson can be easily computed,
and the result for the anomalous form factor has the form:

\begin{equation}
\label{AFF} A\left( q^2, k_1^2, k_2^2\right)= \int\limits_0^1 dx\int
\limits_0^1dy\int\limits_0^1dz\ \frac{2\ \delta(1-x-y-z)}{\left[ 1-
\dfrac{q^2}{m_N^2}\,xy-\dfrac{k_1^2}{m_N^2}\,xz-\dfrac{k_2^2}{m_N^2}\,yz-
i\epsilon \right] }
\end{equation}

\noindent
where $q$, $k_1$ and $k_2$ are the four-momenta of the PGB and the two gauge
bosons, and $m_N$ is the nucleon mass. The form factor is properly normalized
at zero momenta, $A(0,0,0)=1$.

\begin{figure}[t]
\setlength{\epsfxsize}{100mm}\epsfbox[45 240 320 350]{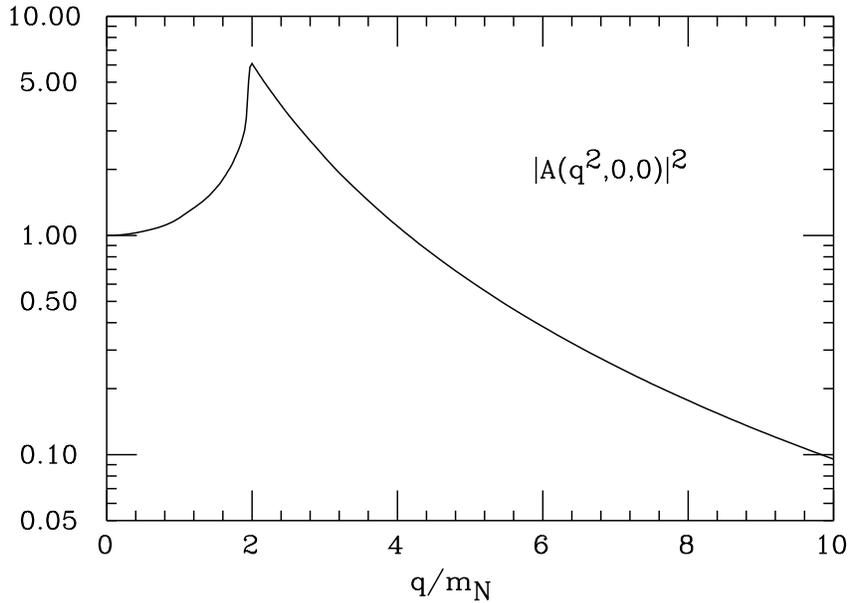}
\vspace{3.0truecm}
\caption[]{\it The form factor $\vert A\left(q^2 , k_1^2, k_2^2\right) \vert
^2$ as a function of the dimensionless variable $\sqrt{q^2/m_N^2}$, for
$k_1^2=k_2^2=0$.}
\protect\label{fig:AFF}
\end{figure}

In fig. \ref{fig:AFF}, $\vert A\left( q^2, k_1^2, k_2^2\right) \vert ^2$ is
shown as a function of the dimensionless variable $\sqrt{q^2/m_N^2}$, for
$k_1^2=k_2^2=0$. This choice corresponds, for instance, to the $P \ra 2\gamma$
decay, with $q^2=M_P^2$, or to the $e^+e^- \ra P \gamma$ process, when $q^2$
is the energy square in the center of mass and $M_P^2=0$.
%___________________________________________________________________________
%___________________________________________________________________________

Fig. \ref{fig:AFF} shows that, in the region of small $q^2$, the form factor
$\vert A\left(q^2,0,0 \right)\vert ^2$ is an increasing function of
the momenta, until it reaches a peak in correspondence of $q^2=4m_N^2$. For
$q^2 \geq 4m_N^2$, the form factor acquires a non zero imaginary part, that
would correspond, according to the Cutkosky rule, to the possible creation of
a real nucleon pair in the $P \ra NN$ channel. Above this threshold, the form
factor starts to decrease for increasing $q^2$, thus correcting the bad
asymptotic behaviour of the corresponding cross sections.

Let us now discuss these results in the framework of QCD. In QCD, the
anomalous $\pi^0 \ra 2\, \gamma$ decay rate, calculated in the limit $m_{\pi}=
0$, is in agreement with the experimental value within less than a few per
cent of accuracy. In this case, the anomalous form factor, as given from eq.
(\ref{AFF}), would only introduce a correction of the order of $0.3\%$, thus
strongly supporting the validity of the $m_{\pi}=0$ approximation.

A priori, one could expect that very large corrections to the couplings of eq.
(\ref{APBB}) would be found in the case of the $\eta \ra 2\,\gamma$ and $\eta
\prime \ra 2\, \gamma$ decays, since the masses of these mesons are of order
of two or three times $\Lqcd$. However, the evaluation of the anomalous form
factor, in the linear $\sigma$ model, indicates that this is not the case. For
a PGB mass equal to the $\eta \prime$ mass, we find that this correction is of
the order of $20\%$. On the other hand, the $\eta \ra 2\,\gamma$ and $\eta
\prime \ra 2\, \gamma$ decay rates, when computed in the limit of zero meson
masses, are consistent, with the corresponding experimental values, within
approximately a factor of 1.5, with a large theoretical uncertainty coming, in
this case, from the unknown value of the $\eta - \eta \prime$ mixing angle.
This uncertainty is too large to draw any definite conclusion. However, the
estimate of the anomalous form factor, obtained in the linear $\sigma$ model,
seems consistent with the experimental data.

These arguments suggest that the high energy corrections, to the anomalous
couplings of eq. (\ref{APBB}), only become relevant, in QCD, when the typical
energy scale in the process is of the order of several times the nucleon mass.
In particular, the anomalous form factor of fig. \ref{fig:AFF}, is found to be
still of the order of one when the energy scale is equal to $4\, m_N$, and it
is reduced by one order of magnitude only when the energy becomes of the order
of $\sim 10 \, m_N$.

Let us now apply these results in the framework of TC theories. The relevant
energy scale, entering in the expression of the form factor, is expected to be
the techninucleon mass (or, possibly, the mass of the lightest technihadron
coupled in pair to the technipion). In a QCD-like TC model, this mass can be
simply estimated by scaling from QCD and using large N arguments. One then
finds the result:

\begin{equation}
\label{MTN} M_{T-Nucleon} = \left(\frac{\Lqcd}{\Ltc}\right)
\left(\frac{N_{TC}}{3}\right) m_{Nucleon}
\end{equation}

\noindent
For a WTC model, the simple scaling from QCD is not expected to be accurate.
However, by not having a better prescription, we can only rely on this
approximation. Thus, when applied for example to the LR model, eq. (\ref{MTN})
indicates that, in the technilepton and techniquark sector, the techninucleon
masses are in the range between 400 and 800 GeV, depending on the set of
parameters, $A$ or $B$ in table \ref{tab:C}, one is considering.

In practice, the precise values of these masses are not relevant for our
purposes. We just observe that, if the technihadrons have masses in the range
of several hundred GeV, then the corrections introduced by the anomalous form
factor to the couplings of eq. (\ref{APBB}) are likely to be of the order of
only $10-20\%$, either for PGB masses of the order of $\Ltc$, or for processes
occurring at the 500 GeV energy of the NLC experiment. On the other hand,
these corrections are likely to be of the same order of magnitude of the
theoretical uncertainties affecting the calculation of the form factor itself.
For these reasons, we will not attempt to take them into account. We just
conclude that, in all the processes we are going to consider in this paper,
the interactions between PGBs and ordinary gauge bosons are expected to be
described, reasonably well, by the couplings of eq. (\ref{APBB}).

\subsection{Couplings of neutral PGBs to ordinary fermions}

The PGBs of TC theories are also coupled to the ordinary fermions by the ETC
interactions. In models of TC, the ETC interactions are introduced in order
to explicitly break the fermion chiral symmetries, and to generate the
ordinary fermion and technifermion masses \cite{DS,EL1}. In the low energy
limit, with respect to $\Letc$, the ETC interactions can be described in terms
of an effective Lagrangian, which contains four-fermion couplings between
ordinary fermions and technifermions. These couplings also describe the
interactions between the PGBs and the ordinary fermion pairs.

The most general form of the ETC effective Lagrangian has been discussed in
ref. \cite{EL1}. It is possible to show, starting from this general case, that
the coupling between a neutral PGB $P$ and an ordinary fermion pair, $\bar
f f$, is expected to be proportional to the fermion mass $m_f$, and to the
inverse of the pseudoscalar decay constant $F_P $, $g_{Pff} \sim m_f/F_P$
\footnote{This is the analog of the Goldberger and Treiman relation in QCD,
which describes the effective coupling between the pion and a nucleon pair.}.

Starting from this result, and following ref. \cite{VL}, we then write the
coupling between a neutral PGB $P$ and a fermion pair $\bar ff$ in terms of
a second class of model dependent constants, $B_{Pff}$, in the form:

\begin{equation}
\label{BPff}-B_{Pff}\,\left( \dfrac{m_f}{F_P\sqrt{n/2}}\right) P\,(\overline{%
f} \,i\,\gamma ^5\,f)
\end{equation}

\noindent
The precise values of the constants $B_{Pff}$ depend on the structure of the
ETC gauge group and the fermion and technifermion contents of the ETC
representations. However, with the definition of eq. (\ref{BPff}), all these
constants are typically expected to be of order one, and we refer the reader
to ref. \cite{VL} for some specific examples.

In this paper, we will not go in further details on this point. By having
defined the general form of the PGB couplings to both ordinary gauge bosons
and fermions, eqs. (\ref{APBB}) and (\ref{BPff}), we now proceed to compute
the various rates of PGB production and decay in the $e^+e^-$ collision
experiments.

\section{The production and decay of neutral PGBs in $e^+e^-$ collisions}

In the analysis of the LR model, performed in section 2, we found that the
masses of the three lightest PGBs are approximately in the range between 100
and 350 GeV (see eqs. (\ref{MP03N}) and (\ref{MPAVN})). One can argue that
this is a quite general result. In WTC models, the masses of those light PGBs,
that are only generated by the ETC interactions, are generally estimated to be
of the order of the relevant TC scale, $\Ltc$. In models with a single scale
of dynamical symmetry breaking, $\Ltc$ is roughly of the order of 500 GeV. In
multiscale models, where a few low energy scales can appear, it is unlikely
that these scales, and, hence, the PGB masses, can be smaller than 100 GeV.
For this reason, in the rest of this paper, by considering the results of eqs.
(\ref{MP03N}) and (\ref{MPAVN}), we will only allow the neutral PGB masses to
vary in the range between 100 and 350 GeV.

{}From a phenomenological point of view, it follows that, in order to discuss
the PGB production in $e^+e^-$ collision, it is reasonable to assume that the
PGBs of WTC are too heavy to be produced from on-shell $Z^0$ decays, e.g. at
the LEP I or SLC experiments. Thus, we will only consider in this paper the
PGB production at the two presently planned high energy $e^+e^-$ colliders,
namely LEP II and NLC. For these experiments, we will assume the energy in the
center of mass to be equal to 200 and 500 GeV, and the integrated luminosity
per year to be equal to $5\cdot 10^2$ and $10^4$ pb$^{-1}$ respectively.

The several cross sections and decay rates, relevant for the PGB production
in $e^{+} e^{-}$ collisions, will be computed in the framework of a generic
TC/ETC model. The dependence of the results on the particular model is
introduced by the value of the PGB decay constant, $F_P$, the number of
technicolors, $N_{TC}$, the dimension $n$ of the flavour symmetry group, and
the PGB couplings, $A_{PB_1B_2}$ and $B_{Pff}$. From eqs. (\ref{APBB}) and
(\ref{BPff}), it is clear that the two model dependent quantities, $F_P$ and
$n$, will always appear through the combination $F_P \sqrt{n/2}$. In
traditional single-scale TC models this combination is constrained to be equal
to the electroweak scale, $F_\pi=246$ GeV. Thus, as far as the traditional TC
models are concerned, our predictions are actually independent on the value of
the specific TC scale.

In presenting the numerical results for WTC models we will consider, in
particular, the predictions obtained in the framework of the LR model
discussed in section 2. In this case, in order to simplify the discussion, we
will identify the mass eigenstates, in the light PGB sector, with the three
weak isospin eigenstates defined in eq. (\ref{P30LR}), namely the PGBs $P^0$,
$P^3_Q$ and $P^3_L$. Thus, as far as these PGBs are concerned, we will neglect
the mixing introduced in the model by the explicit isospin symmetry breaking.
With this assumption, the values of the couplings $A_{PB_1B_2}$, between the
PGBs and the ordinary gauge bosons, are those given in table \ref{tab:B}. To
be specific, we will only consider, in the following, the values of technipion
decay constants labelled as set $A$ in table \ref{tab:C}. With the values
denoted as set $B$, we would have obtained smaller values, of the order of
$50\%$, for the corresponding cross sections (this ratio scales approximately
as $(F^A/F^B)^2$).

As already observed in section 2, one of the unrealistic aspects of the LR
model is the flavour content of ordinary fermions. With $N_L=6$, the model
predicts the existence of six doublets of quarks, one doublets of antiquarks
and $N_L\,(N_L-1)\,/\,2=15$ doublets of ordinary leptons. On the other hand,
the choice of a smaller value of $N_L$ (e.g. the more ``realistic" choice
$N_L=3$) would not guarantee anymore the walking of the TC coupling constant
and, consequently, the onset of WTC dynamics. Thus, in order to obtain
sensible predictions, we will keep $N_L$ fixed to the value $N_L=6$, and we
will simply assume that, among the whole set of ordinary fermions, there exist
the three standard families of quarks and leptons observed in the experiments.
Only these particles will be then considered as possible candidates in the
final states. Furthermore, in presenting all our numerical results, we will
allow the values of the model dependent ETC couplings, $B_{PB_1B_2}$, to vary
in the range between 1/3 and 3.

\subsection{The decays of PGBs in multiscale WTC models}

In order to investigate the experimental signatures expected when the PGBs are
produced in $e^+e^-$ collisions, we first consider, in this section, the
possible decay modes of these particles. Since, within the framework of
traditional TC models, the PGB decays have been already extensively discussed
in the literature (see e.g. refs. \cite{EHLQ} and \cite{EGNS}), we will
concentrate here on those aspects of neutral PGB decays that are peculiar of
WTC dynamics.

By being the lightest states among all the existing technihadrons, the neutral
PGBs we are considering in this paper are not allowed to decay via the strong
TC interactions. Thus, they can only decay either into a pair of standard
gauge boson, through the anomalous couplings of eq. (\ref{APBB}), or into an
ordinary fermion-antifermion pair, through the ETC couplings of eq.
(\ref{BPff}). The corresponding decay widths, expressed in terms of the model
dependent constants $A_{PB_1B_2}$ and $B_{Pff}$, are given by:

\begin{equation}
\label{PBBd} \Gamma (P\ra B_1B_2)=
\dfrac{N_C}{(1+\delta _{B_1B_2})} \ \,\left( \dfrac{\alpha^2 A_{PB_1B_2}^2
N_{TC}^2} {32\pi ^3F_P^2\,(n/2)}\right) \ M_P^3\ \lambda
\left( 1,\,\dfrac{M_{B_1}^2}{M_P^2},\,\,\dfrac{M_{B_2}^2}{M_P^2}\right)
\end{equation}

\noindent and

\begin{equation}
\label{Pffd} \Gamma (P\ra \bar ff)=
N_C\,\left( \dfrac{B_{Pff}^2}{8\pi F_P^2\,(n/2)}\right) \,M_P\,m_f^2\,
\sqrt{ 1-\dfrac{4m_f^2}{M_P^2}}
\end{equation}

\noindent
In these equations, $N_C$ represents the number of colors of the two particles
in the final state: $N_C=3$ for quarks, $N_C=8$ for gluons and $N_C=1$ for
leptons and electroweak gauge bosons. In addition, the function $\lambda$ is
defined as:

\begin{equation}
\label{lambda} \lambda (a,b,c)\equiv a^2+b^2+c^2-2\,ab-2\,bc-2\,ac
\end{equation}

Since the decay widths in eqs. (\ref{PBBd}) and (\ref{Pffd}) are both
proportional to $1/F_P^2$, the corresponding branching ratios turn out to be
independent on the value of the specific TC scale. Thus, we find that the
existence of low energy scales, in multiscale WTC models, does not affect the
relative weight of the various branching ratios, with respect to the
predictions of traditional TC models. The only effect of such low scales, is a
net increasing of the total PGB decay widths.

In contrast, a considerable effect of WTC dynamics in PGB decays, comes from
the typically large values of PGB masses. Indeed, the decay widths of PGBs
into gauge boson pairs are proportional to the third power of the PGB mass,
while the decay widths into ordinary fermion pairs only increase linearly with
this mass. Thus, we expect that the large PGB masses, predicted in WTC models,
will enhance the several $P\ra B_1B_2$ branching ratios with respect to those
of the $P\ra \bar ff$ decays.

As an example, let us consider the partial decay widths of the three light
neutral PGBs of the LR model. The two isovectors, $P^3_Q$ and $P^3_L$, are not
coupled to a QCD gluon pair. Thus, the gluon-gluon decay is only allowed for
the PGB $P^0$. In this case, the relevant coupling, $A_{P^0g^ag^a}$, is equal
to $\sqrt{2}$, and one finds that, for this particle, the decay mode into a
gluon pair represents the favored channel. For instance, by assuming that the
ETC coupling constants, $B_{Pff}$, are all equal to one, we find that the $P^0
\ra gg$ branching ratio can vary between $92\%$, for $M_P=100$ GeV, and $99\%$,
for $M_P=350$ GeV. If we consider larger values of the ETC couplings, for
example $B_{Pff}=3$, then we find that the PGB $P^0$ is also expected to decay
in a $\bar bb$ pair in approximately $40\%$ of the cases, if its mass is equal
to 100 GeV. However, this probability is reduced to less than $15\%$ for
$M_P=200$ GeV, and to approximately $5\%$ for $M_P=350$ GeV.

If the PGBs are not coupled to a gluon pair, as the PGBs $P^3_Q$ and $P^3_L$
of the LR model, then the preferred decay mode of these particles is expected
to be into a pair of bottom quarks. For the PGB $P^3_Q$, for example, if the
ETC constants $B_{Pff}$ are all equal to one, the $P^3_Q\ra \bar bb$ branching
ratio varies between $87\%$ and $78\%$, as its mass increases from 100 to 350
GeV. The remaining decay modes occur mainly into a pair of $\bar cc$ ($\simeq
7-8\%$) or $\tau^+\tau^-$ ($\simeq 3-4\%$). However, for smaller values of the
ETC couplings, and for large values of the PGB mass, a considerable fraction
of the PGB decays can also occur into the anomalous channel $P \ra\gamma
\gamma$. For $B_{Pff}=1/3$ and $M_P=200$ GeV, the $P\ra \gamma\gamma$
branching ratio is equal to approximately $26\%$. For even larger PGB masses,
for example $M_P=350$ GeV, the decay mode into a photon pair is the dominant
one, with a branching ratio of approximately $50\%$. In this case, the
fraction of $P\ra \bar bb$ decay mode is of the order of $40\%$.

\begin{figure}[t]
\ewxy{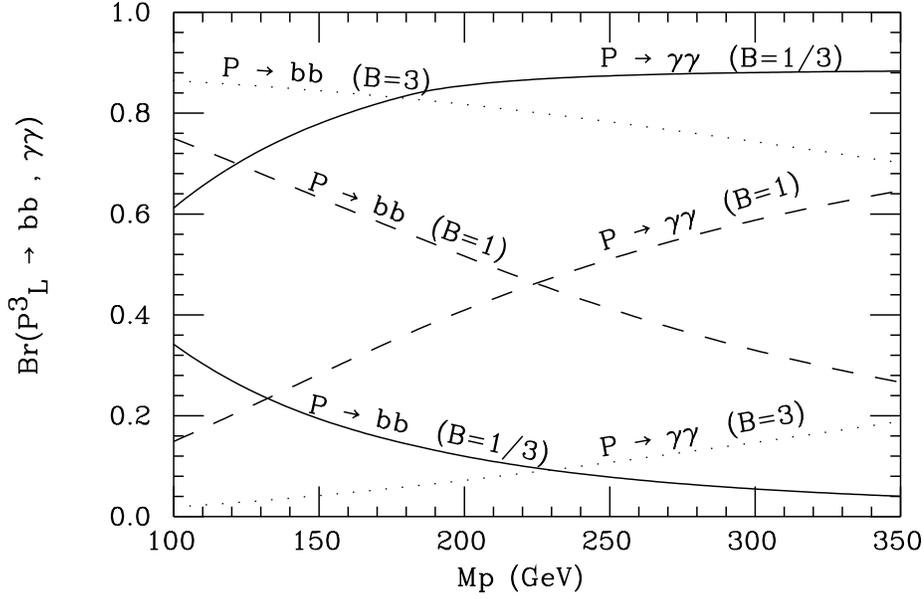}{100mm}
\vspace{3.0truecm}
\caption[]{\it The $P\ra \gamma\gamma$ and $P\ra \bar bb$ branching ratios,
for the PGB $P^3_L$ of the LR model, as a function of the PGB mass $M_P$. The
curves correspond to the values $B_{Pff}=1/3$ (solid), $B_{Pff}=1$ (dashes)
and $B_{Pff}=3$ (dots) of the ETC coupling constants.}
\protect\label{fig:P3Ldec}
\end{figure}

The decay mode into a photon pair is even more relevant for the PGB $P^3_L$.
In this case, in fact, the corresponding coupling $A_{P_L^3\gamma \gamma }$
has a value which is significantly larger with respect to one: $A_{P_L^3\gamma
\gamma }\simeq -7.35$ (see table \ref{tab:B}). The resulting scenario is then
illustrated in fig. \ref{fig:P3Ldec}, in which the values of the $P\ra \gamma
\gamma$ and $P\ra \bar bb$ branching ratios, for the PGB $P^3_L$, are plotted
as a function of the PGB mass. In the figure, the three values $B_{Pff}=1/3$,
1 and 3 have been considered.
%___________________________________________________________________________
%___________________________________________________________________________
By looking at fig. \ref{fig:P3Ldec}, we find that, when $B_{Pff}$ is equal to
1/3, the $P^3_L \ra \gamma\gamma$ channel represents the dominant decay mode
of this particle, regardless of the value of its mass. The corresponding
branching ratios vary between $61\%$, for $M_P=100$ GeV, and $88\%$, for
$M_P=350$ GeV. For $B_{Pff}=1$, the decay mode into a photon pair can still be
dominant, but only for a PGB mass larger than approximately 220 GeV. Finally,
for an even larger value of the ETC coupling, $B_{Pff}=3$, the PGB $P^3_L$ is
expected to decay mainly into a $\bar bb$ pair.

In conclusion, we see that, in WTC theories, the predictions for PGB decays
can significantly differ from those obtained in the framework of traditional
TC models. Due to the large values of masses expected in WTC theories, the
$P \ra \gamma\gamma$ decay can represent a large fraction of
the all PGB decay modes, whereas this channel is usually strongly suppressed
in traditional models. The specific values of PGB branching ratios, however,
are to a large extent model dependent. They strongly depend on the values of
the PGB masses and the values of the model dependent coupling constants.
Nevertheless, in performing a phenomenological analysis, we can take advantage
of the fact that only few relevant decay modes of these particles have to be
considered, namely the decays $P\ra gg$, $P\ra \bar bb$ and $P\ra\gamma
\gamma$. In addition, the $P\ra gg$ decay is forbidden for all the PGBs with
isospin one.

\subsection{The production of PGBs in the $e^{+}e^{-} \ra P\gamma $ and
$e^{+}e^{-} \ra PZ^0$ channels}

We now consider the production of PGBs in $e^+e^-$ collisions. In these
experiments, the simplest channels of neutral PGB production are represented
by the processes:

\begin{equation}
\label{eePB} e^{+}e^{-}\ra P\gamma \qquad , \qquad e^{+}e^{-}\ra PZ^0
\end{equation}

\begin{figure}[t]  % produce figure here
\setlength{\epsfxsize}{100mm}\epsfbox[90 340 370 450]{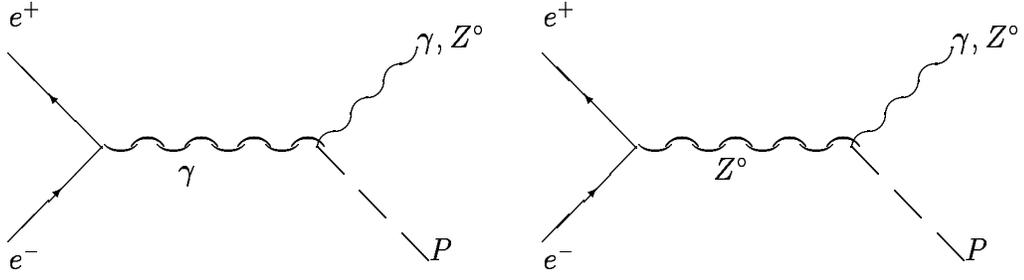}
\vspace{1.0truecm}
\caption[]{\it Feynman diagrams relevant for the processes $e^+e^- \ra P
\gamma$ and $e^+e^- \ra P Z^0$.}
\protect\label{fig:eePBfd}
\end{figure}

\noindent
At the lowest order of perturbation theory, each of these processes is
described by the two Feynman diagrams of fig. \ref{fig:eePBfd}. Both these
diagrams are only controlled by the anomalous TC couplings between the PGBs
and the ordinary gauge bosons, and they do not involve the presence of the ETC
interactions.
%___________________________________________________________________
%___________________________________________________________________

Since the lightest PGB masses are likely in the range between 100 and 350 GeV,
then the $e^{+}e^{-}\ra PZ^0$ channel is practically forbidden at the LEP II
experiment, while the $e^{+}e^{-}\ra P\gamma $ is allowed, provided that the
PGBs have masses smaller than 200 GeV. In the NLC experiment, with an energy
of 500 GeV in the center of mass, both these channels are open.

The total cross section, for the processes in eq. (\ref{eePB}), has been
computed in ref. \cite{Ran2}, where, however, it is only discussed in the
framework of traditional TC models. In terms of the couplings $A_{PB_1B_2}$,
the result has the form:

\begin{equation}
\label{SeePB}
\renewcommand{\arraystretch}{2.0}
\begin{array}{rl}
\sigma (e^{+}e^{-} & \ra PB) =\left(\dfrac{\alpha ^3\,N_{TC}^2}{24\,\pi ^2\,
F_P^2\,(n/2)}\right) \ \lambda ^{3/2}(1,m_P^2,m_B^2) \,\cdot \\
& \cdot \ \left[ A_{PB\gamma }^2 \,+\,\dfrac{A_{PB\gamma}\, A_{PBZ}
\,(1-4\,s_W^2)}{2\,s_W\,c_W\,(1-m_Z^2)}\,+\, \dfrac{A_{PBZ}^2 \,(1-4\,
s_W^2+8\,s_W^4)}{8\,s_W^2\,c_W^2\,(1-m_Z^2)^2}\right]
\end{array}
\renewcommand{\arraystretch}{1.0}
\end{equation}

\noindent
where $B$ may be a photon or a $Z^0$ gauge boson. In eq. (\ref{SeePB}),
$\sqrt{s}$ is the energy in the center of mass, $m_P$ and $m_B$ are the
particle masses in units of $\sqrt{s}$, and the function $\lambda $ is defined
in eq. (\ref{lambda}). Of the three terms entering in eq. (\ref{SeePB}), the
first and the last one describe the contribution of the two diagrams of fig.
\ref{fig:eePBfd}, while the second term represents the interference between
them. Note that, as discussed in section 3, without the introduction of the
anomalous form factor, the cross section in eq. (\ref{SeePB}), in the high
energy limit, $s\ra\infty$, tends to be constant.

The $\sigma (e^{+}e^{-} \ra PB)$ cross section is proportional to the factor
$1/F_P^2\,(n/2)$. This is the origin of the enhancement of the cross section
in multiscale WTC models. In the LR model, for example, the scale $F_P=F_Q
\simeq F_L$ is of the order of approximately 30 GeV (see table \ref{tab:C}),
and the number of technifermion species is $n=2\,(N_L+3)=18$. Thus, the
quantity $F_P\sqrt{n/2}$ is equal to approximately 90 GeV, and it is smaller
by almost a factor of three with respect to the value $F_P\sqrt{n/2}\simeq
246$ GeV occurring in models with a single TC scale. This means that the
values of the corresponding cross sections are larger by approximately one
order of magnitude.

\begin{figure}[t]
\ewxy{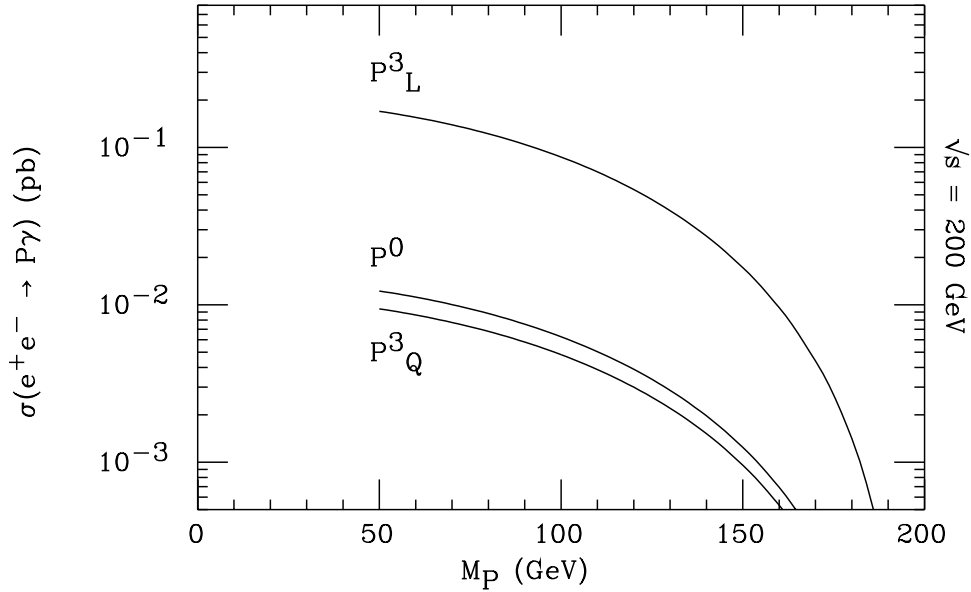}{100mm}
\vspace{3.0truecm}
\caption[]{\it The $\sigma (e^{+}e^{-}\ra P\gamma )$ cross sections, in the LR
model, as a function of the PGB mass $M_P$. The energy in the center of mass
is fixed to the value of $200$ GeV.}
\protect\label{fig:eePg2}
\end{figure}

Eq. (\ref{SeePB}) can be used to compute the $\sigma (e^{+}e^{-}\ra P B)$
cross section in any model of interest. In order to discuss these processes
in a more quantitative way, we show in fig. \ref{fig:eePg2} the values of
this cross section, in the $e^{+}e^{-}\ra P\gamma$ channel, as a function of
the PGB mass $M_P$, for the three PGBs $P_L^3$, $P_Q^3$ and $P^0$ of the LR
model. The energy in center of mass is fixed to the LEP II value of 200 GeV.
%___________________________________________________________________________
%___________________________________________________________________________
We see from the figure that the total cross section, in the case of the PGB
$P_L^3$, is larger by approximately one order of magnitude with respect to the
values obtained for the PGBs $P_Q^3$ and $P^0$. The reason of such a
difference is the large value that the coupling $A_{P_L^3\gamma \gamma }$
assumes in the LR model. For this particle, we find that the values of the
$\sigma (e^{+}e^{-}\ra P_L^3 \gamma )$ cross section, at the energy of 200
GeV, vary between $10^{-1}$ and $10^{-2}$ pb, in the range of PGB masses
between 100 and 150 GeV. This corresponds, at LEP II, to a production rate of
approximately 10-50 PGBs per year. Due to the strong phase space suppression,
this rate becomes then negligible when the PGB mass is larger than 150 GeV.

On the other hand, when we consider the PGBs $P_Q^3$ and $P^0$, we find from
fig. \ref{fig:eePg2} that the number of $e^{+}e^{-}\ra P\gamma$ events per
year is expected to be only of the order of few units, even for PGB masses of
approximately 100 GeV. For the two PGBs, $P_Q^3$ and $P^0$, the anomalous
couplings to the photons and $Z^0$ gauge bosons are quite close to their
typical values of one. Therefore, a cross section of the order of $10^{-2}$
pb is a typical value expected in multiscale WTC models. We then conclude
that, in order the PGBs can be produced at the LEP II experiment, in this
channel, the corresponding $A_{PB_1B_2}$ coupling constants must be, in some
of the relevant cases, larger than one. This can happen, and, in particular,
this is just the case of the LR model of WTC.

\begin{figure}[t]
\ewxy{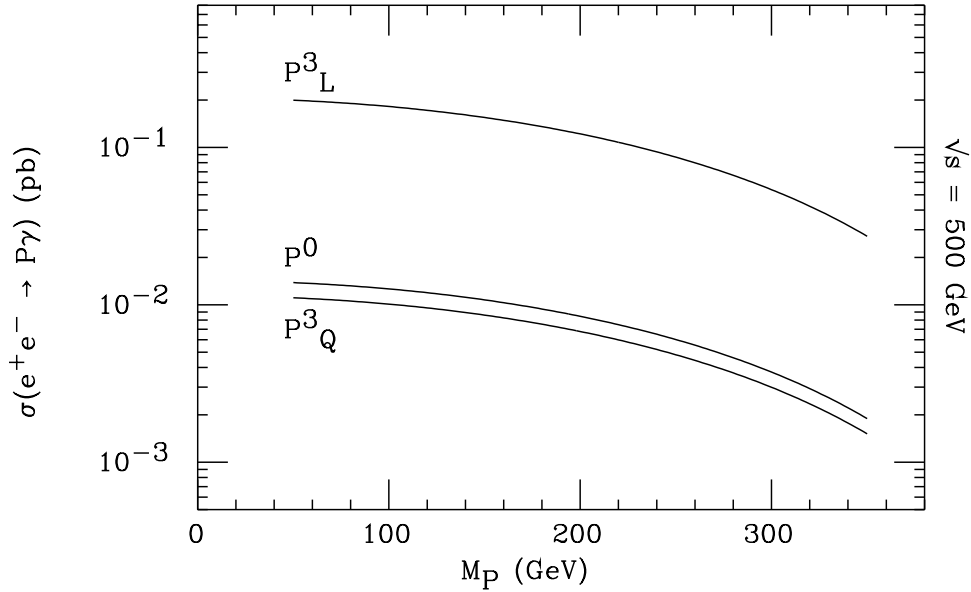}{100mm}
\vspace{3.0truecm}
\caption[]{\it The $\sigma (e^{+}e^{-}\ra P\gamma )$ cross sections, in the LR
model, as a function of the PGB mass $M_P$. The energy in the center of mass
is fixed to the value of $500$ GeV.}
\protect\label{fig:eePg5}
\end{figure}

The rate of PGB production, for the processes of eq. (\ref{eePB}), is expected
to be significantly larger at a 500 GeV NLC experiment. For the three PGBs of
the LR model, the values of the $\sigma (e^{+}e^{-}\ra P\gamma )$ cross
section are shown in fig. \ref{fig:eePg5}, as a function of the PGB mass $M_P$.
%___________________________________________________________________________
%___________________________________________________________________________
In the case of the PGB $P_L^3$, the cross section is of the order of $10^{-1}$
pb, in the whole range of masses typically expected for this particle. By
assuming a luminosity of $10^4$ pb$^{-1}$/yr, this means that the number of
PGBs produced per year is approximately $10^3$. In the case of the PGBs
$P_Q^3$ and $P^0$ this number is then reduced by approximately one order of
magnitude.

In the NLC experiment, the PGB production can also occur in the $e^{+}e^{-}\ra
PZ^0$ channel. Typically, in multiscale WTC models, the values of the
corresponding cross sections are found to be between $10^{-2}$ and $10^{-3}$
pb, so that a considerable rate of PGB production can be expected also in this
channel.

To make a comparison, we now consider the $e^{+}e^{-}\ra PB$ processes in the
framework of traditional, single scale, TC models. We remind that in this case
the quantity $F_P\sqrt{n/2}$ is equal to 246 GeV, so that the only model
dependence of the cross section in eq. (\ref{SeePB}) comes from the number of
technicolors $N_{TC}$ and the values of the $A_{PB_1B_2}$ coupling constants.
In the one family FS model, for instance, all these constants are quite close
to their typical value of one (see table \ref{tab:B}). In this case, even by
assuming the large value $N_{TC}=8$ and by considering PGB masses of the order
of 50 GeV, we find that the total number of $e^{+}e^{-}\ra P\gamma$ and $e^{+}
e^{-}\ra PZ$ events is expected to be approximately of only 10 per year at
NLC, whereas no PGBs will be typically produced, in these channels, at the LEP
II experiment.

\subsection{Analysis of the $e^{+}e^{-}\ra P\gamma$ events at LEP II}

Let us now discuss, in some detail, the experimental signatures of the $e^+e^-
\ra P \gamma$ events and the predicted Standard Model background. To be
specific, we will consider in this section the values of model parameters and
couplings of the PGB $P_L^3$ of the LR model. In practice, the precise values
of these couplings are not relevant for the following analysis, whose results
can be applied, at least qualitatively, to any other model of WTC. In this
respect, the only relevant assumption is that the model contains one PGB whose
anomalous coupling with a photon pair is large enough for this particle to be
produced at LEP II.

Let us first assume that the mass of the PGB is fixed to the value of 100 GeV.
In this case, the main results of the analysis we are going to perform are
summarized in table \ref{tab:C1}.
%_________________________________________________________________________
\begin{table} \centering
\renewcommand{\arraystretch}{1.5}
\begin{tabular}{|c||c|c|c|} \hline
{\bf M$_P$ = 100 GeV}
& $B_{Pff} = 1/3 $ & $B_{Pff} = 1 $ & $B_{Pff} = 3 $ \\ \hline \hline
$\sigma (e^+e^- \ra P \, \gamma)$
&$ 0.87 \, \cdot \, 10^{-1} $
&$ 0.87 \, \cdot \, 10^{-1} $
&$ 0.87 \, \cdot \, 10^{-1} $ \\ \hline
$Br(P \ra \bar bb) $
&$ 0.34 $  &$ 0.75 $  &$ 0.86 $ \\
$Br(P \ra \gamma \gamma) $
&$ 0.61 $  &$ 0.15 $  &$ 0.02 $  \\ \hline
$\sigma_S(e^+e^- \ra \bar bb \gamma)_{cut} $
&$ 0.27 \, \cdot \, 10^{-1} $
&$ 0.59 \, \cdot \, 10^{-1} $
&$ 0.68 \, \cdot \, 10^{-1} $ \\
$N_S(e^+e^- \ra \bar bb \gamma) $
&$ 14 $  &$ 30 $  &$ 34 $ \\
$N_B(e^+e^- \ra \bar bb \gamma) $
&$ 10 $  &$ 10 $  &$ 10 $ \\ \hline
$\sigma_S(e^+e^- \ra \gamma \gamma \gamma)_{cut} $
&$ 0.43 \, \cdot \, 10^{-1} $
&$ 0.10 \, \cdot \, 10^{-1} $
&$ 0.01 \, \cdot \, 10^{-1} $ \\
$N_S(e^+e^- \ra \gamma \gamma \gamma) $
&$ 21 $  &$ 5 $  &$ 1 $  \\
$N_B(e^+e^- \ra \gamma \gamma \gamma) $
&$ 6  $  &$ 6 $  &$ 6 $  \\ \hline
\end{tabular}
\renewcommand{\arraystretch}{1.0}
\caption{\it Values of the cross sections (in pb), branching ratios and
expected number of events per year for the $e^+e^- \ra P^3_L \gamma$ processes
and the relevant Standard Model background at the LEP II experiment. $N_S$ and
$N_B$ refer to the signal and background respectively. For the ``cut" see text.
The PGB mass is fixed to the value $M_P=100$ GeV.
\label{tab:C1}}
\end{table}
%_________________________________________________________________________

At LEP II, the $\sigma(e^+e^- \ra P^3_L \gamma)$ cross section, for the PGB
$P^3_L$ with a mass of 100 GeV, is equal to approximately $0.87\cdot 10^{-1}$
pb. This corresponds to a production rate of more than 40 PGBs per year. The
$P^3_L$ is then expected to decay mainly into a $\bar bb$ or into a $\gamma
\gamma$ pair. As the ETC coupling constants, $B_{Pff}$, increase from 1/3 to
3, the corresponding branching ratios vary between $34\%$ and $86\%$ in the
$\bar bb$ channel, and between $61\%$ and $2\%$ for the decay into a photon
pair.

If the PGB decays into a pair of bottom quarks, then the expected signal is an
event with two high energy, well isolated, jets and a monochromatic photon.

Up to small corrections, of $O(m_b^2/M_P^2)$, the minimum energy of each jet
is equal to $M_P^2/4E=25$ GeV, where E=100 GeV is the energy of the electron
beam. In addition, the two jets are separated by an angle, $\theta_{bb}$, that
is always larger than the minimum value $\theta_{bb}^{\rm min}=$ arccos$\,(2
\beta_P^2-1)$, where $\beta_P=(4E^2-M_P^2)/(4E^2+M_P^2)$ is the speed of the
PGB. In the case we are considering, $\beta_P=0.6$ and $\theta_{bb}^{\rm min}
\simeq 106^{\rm o}$.

The main signature of these events is the presence of a monochromatic photon,
whose energy is given by:

\begin{equation}
\label{Egamma} E_{\gamma} = E - \frac{M_P^2}{4E} = 75 \ {\rm GeV}
\end{equation}

\noindent
Thus, the behaviour of the differential cross section, $d\sigma/dE_{\gamma}$,
is represented, for the signal, by a sharp peak in correspondence of this
energy.

The finite width of the peak is determined by three effects. The uncertainty
on the beam energy, the experimental error in the measure of the photon energy
and the physical width of the PGB. In this analysis we assume that, at the LEP
II experiment, the uncertainty on the beam energy will be equal to $\Delta E=
30$ MeV, and the accuracy in measuring the photon energy will be $\Delta E_
{\gamma}/E_{\gamma}=1.2\%$. As far as the PGB width is concerned, its value
can vary considerably with the values of the ETC constants $B_{Pff}$. If these
constants are assumed to be all equal, then we find that $\Gamma_P$ can vary
between 13 MeV, for $B_{Pff}=1/3$, and 420 MeV, for $B_{Pff}=3$. Thus, the
ratio $\Gamma_P/M_P$ is expected to be at most of the order of $4\cdot 10^
{-3}$. We then find that the width of the peak, in the shape of $d\sigma/dE_
{\gamma}$, turns out to be mainly determined by the experimental error in the
measure of the photon energy, and it is expected to be equal to approximately
1 GeV.

Within the Standard Model, the total $e^+e^- \ra \bar qq \gamma$ cross section
has been computed, to the lowest order in perturbation theory, in ref.
\cite{GR}. This cross section turns out to be divergent when the final photon
becomes collinear with the beam direction. Thus, in order to compute the
expected background, we have applied the cut $20^{\rm o} \le \theta_{\gamma e}
\le 160^{\rm o}$ on the angle between the photon and the initial electrons. In
addition, according to eq. (\ref{Egamma}) and the estimated value of $\Gamma
_P$, we have only considered photons with energy between 74 and 76 GeV. In
this way, we find that, in the Standard Model, the $\sigma(e^+e^- \ra \bar bb
\gamma)$ background cross section, for $\sqrt{s}=200$ GeV, is equal to
approximately $2 \cdot 10^{-2}$ pb. This corresponds, at LEP II, to 10 events
per year.

The differential cross section, $d\sigma/d\cos \theta_{\gamma e}$, for the
$e^+e^- \ra P^3_L \gamma$ events, is proportional to $1+ \cos^2 \theta_{\gamma
e}$. Thus, the above cuts on the angle $\theta_{\gamma e}$ implies the loss of
approximately $9\%$ of the signal. We then find that the number of PGB
expected per year in this channel varies from 14, for $B_{Pff}=1/3$, to 34,
for $B_{Pff}=3$ (see table \ref{tab:C1}). The ratio signal/background is
therefore equal to 1.4, when $B_{Pff}=1/3$, and it is always larger than 3 if
$B_{Pff}$ is larger than 1. Thus, these events are expected to be clearly
distinct from the Standard Model background.

In order to estimate the number of events that can be effectively observed at
LEP II, we must finally consider the experimental efficiency in detecting $b$
quarks. In these events, at least one of the two final $b$ quarks must be
reconstructed, and the corresponding efficiency can be assumed to be of the
order of $65\%$, if the purity of the sample is approximately $85\%$. Thus, in
the case of the PGB $P_L^3$, we find that more than 20 events per year can be
observed at LEP II, in the $e^+e^- \ra P^3_L \gamma \ra \bar bb \gamma$
channel, provided the ETC coupling constants are equal or larger than one.

If the ETC coupling constants are smaller than one, then the PGB $P^3_L$ is
expected to decay mainly into a photon pair. For $M_P=100$ GeV and $B_{Pff}=
1/3$, the $P^3_L \ra \gamma\gamma$ branching ratio is equal to $61\%$. In this
case, the WTC event is characterized by the spectacular signature of only
three photons in the final state. One of these photons has energy that is
still given by eq. (\ref{Egamma}), and the other two have energies larger than
25 GeV.

In the Standard Model, the $\sigma(e^+e^- \ra 3\gamma)$ cross section has been
calculated in ref. \cite{BKCGT}, in the limit of vanishing electron mass.
By using their results, we have computed the value of this cross section by
requiring at least one photon with an energy between 74 and 76 GeV, and the
two other photons with energies larger than 25 GeV. In order to avoid
collinear divergences, we have also required the angles between the photons
and the electron beam to be always in the range between 20 and 160 degrees.
With these cuts, we find that, in the Standard Model, the $\sigma(e^+e^-\ra 3
\gamma)$ background cross section is equal to approximately $1.2\cdot 10^{-2}$
pb, corresponding, at LEP II, to 6 events per year.

By performing a simple Monte Carlo simulation, we have estimated that the
above cuts on the angles between photons and the initial beam reduce the
expected WTC signal by approximately $20\%$. We then find that, for $M_P=100$
GeV and $B_{Pff}=1/3$, the number of produced PGBs, in the $e^+e^-\ra P \gamma
\ra 3 \gamma$ channel, is of the order of 20 per year, and the ratio
signal/background is larger than 3.5. Thus, even in this channel, the WTC
events are expected to be clearly distinct from the Standard Model background.

The above analysis can be repeated by considering different values of the PGB
mass. The results, for $M_P=125$ GeV and $M_P=150$ GeV, are presented in
tables \ref{tab:C2} and \ref{tab:C3} respectively. We find that the ratio
signal/background always increases for increasing PGB masses. However, if
these masses are equal or larger than 150 GeV, the expected number of WTC
events, in each channel, is likely to be too small for these processes to be
observed at LEP II.
%_________________________________________________________________________
\begin{table} \centering
\renewcommand{\arraystretch}{1.5}
\begin{tabular}{|c||c|c|c|} \hline
{\bf M$_P$ = 125 GeV}
& $B_{Pff} = 1/3 $ & $B_{Pff} = 1 $ & $B_{Pff} = 3 $ \\ \hline \hline
$\sigma (e^+e^- \ra P \, \gamma)$
&$ 0.47 \, \cdot \, 10^{-1} $
&$ 0.47 \, \cdot \, 10^{-1} $
&$ 0.47 \, \cdot \, 10^{-1} $ \\ \hline
$Br(P \ra \bar bb) $
&$ 0.25 $  &$ 0.69 $  &$ 0.86 $ \\
$Br(P \ra \gamma \gamma) $
&$ 0.71 $  &$ 0.21 $  &$ 0.03 $  \\ \hline
$\sigma_S(e^+e^- \ra \bar bb \gamma)_{cut} $
&$ 0.11 \, \cdot \, 10^{-1} $
&$ 0.29 \, \cdot \, 10^{-1} $
&$ 0.36 \, \cdot \, 10^{-1} $ \\
$N_S(e^+e^- \ra \bar bb \gamma) $
&$ 5 $  &$ 15 $  &$ 18 $ \\
$N_B(e^+e^- \ra \bar bb \gamma) $
&$ 1 $  &$ 1  $  &$ 1  $ \\ \hline
$\sigma_S(e^+e^- \ra \gamma \gamma \gamma)_{cut} $
&$ 0.27 \, \cdot \, 10^{-1} $
&$ 0.08 \, \cdot \, 10^{-1} $
&$ 0.01 \, \cdot \, 10^{-1} $ \\
$N_S(e^+e^- \ra \gamma \gamma \gamma) $
&$ 13 $  &$ 4 $  &$ 1 $  \\
$N_B(e^+e^- \ra \gamma \gamma \gamma) $
&$ 3  $  &$ 3 $  &$ 3 $  \\ \hline
\end{tabular}
\renewcommand{\arraystretch}{1.0}
\caption{\it Same as in table $4$, but with the PGB mass fixed to the value
$M_P=125$ GeV.
\label{tab:C2}}
\end{table}
%_________________________________________________________________________
%_________________________________________________________________________
\begin{table} \centering
\renewcommand{\arraystretch}{1.5}
\begin{tabular}{|c||c|c|c|} \hline
{\bf M$_P$ = 150 GeV}
& $B_{Pff} = 1/3 $ & $B_{Pff} = 1 $ & $B_{Pff} = 3  $ \\ \hline \hline
$\sigma (e^+e^- \ra P \, \gamma)$
&$ 0.17 \, \cdot \, 10^{-1} $
&$ 0.17 \, \cdot \, 10^{-1} $
&$ 0.17 \, \cdot \, 10^{-1} $ \\ \hline
$Br(P \ra \bar bb) $
&$ 0.19 $  &$ 0.63 $  &$ 0.84 $ \\
$Br(P \ra \gamma \gamma) $
&$ 0.78 $  &$ 0.28 $  &$ 0.04 $  \\ \hline
$\sigma_S(e^+e^- \ra \bar bb \gamma)_{cut} $
&$ 0.03 \, \cdot \, 10^{-1} $
&$ 0.10 \, \cdot \, 10^{-1} $
&$ 0.13 \, \cdot \, 10^{-1} $ \\
$N_S(e^+e^- \ra \bar bb \gamma) $
&$ 2 $  &$ 5 $  &$ 7 $ \\
$N_B(e^+e^- \ra \bar bb \gamma) $
&$ 1 $  &$ 1 $  &$ 1 $ \\ \hline
$\sigma_S(e^+e^- \ra \gamma \gamma \gamma)_{cut} $
&$ 0.11 \, \cdot \, 10^{-1} $
&$ 0.04 \, \cdot \, 10^{-1} $
&$ 0.00 \, \cdot \, 10^{-1} $ \\
$N_S(e^+e^- \ra \gamma \gamma \gamma) $
&$ 5 $  &$ 2 $  &$ 0 $  \\
$N_B(e^+e^- \ra \gamma \gamma \gamma) $
&$ 3 $  &$ 3 $  &$ 3 $  \\ \hline
\end{tabular}
\renewcommand{\arraystretch}{1.0}
\caption{\it Same as in table $4$, but with the PGB mass fixed to the value
$M_P=150$ GeV.
\label{tab:C3}}
\end{table}
%_________________________________________________________________________

\subsection{The production of PGBs in the $e^{+}e^{-}\ra P\bar ff$ channels}

The simple channels of PGB production, discussed in the previous section, are
only controlled by the TC anomalous interactions, that couple the PGBs with
the ordinary gauge bosons. In ref. \cite{VL}, it has been shown that, in the
framework of traditional TC models, the total cross section of PGB production
at the LEP I experiment, although small, receives however a significative
contribution from the ETC interactions, through the effective couplings
between the PGBs and the ordinary fermions. Thus, it is also interesting to
consider the role that the ETC interactions can play at the LEP II and NLC
experiments.

In addition, in $e^+e^-$ collisions occurring at the 500 GeV energy of NLC, is
kinematically allowed the production of top quark pairs, since the top quark
mass is likely of the order of $m_t=174$ GeV \cite{Top}. In this paper we do
not address the question of how such a large quark mass can be generated by
the ETC interactions \cite{CCL}. In order to compute the corresponding cross
sections, we will simply assume that the effective coupling between the PGBs
and the top quark pairs is of the standard form, i.e. $m_t/F_P$. In this case,
in multiscale WTC models, this coupling is typically larger than one, and, a
priori, it can play an important role in PGB production.

In order to investigate the role of the ETC interactions at the LEP II and
NLC energies, we study the processes:

\begin{equation}
\label{eePff} e^{+}e^{-}\ra P\bar ff
\end{equation}

\noindent
in which $\bar ff$ is a fermion pair. The particular case of quasi-elastic
scattering, $e^{+}e^{-}\ra P e^{+}e^{-}$, will be discussed separately in the
following section.

\begin{figure}[t]  % produce figure here
\setlength{\epsfxsize}{100mm}\epsfbox[90 440 370 540]{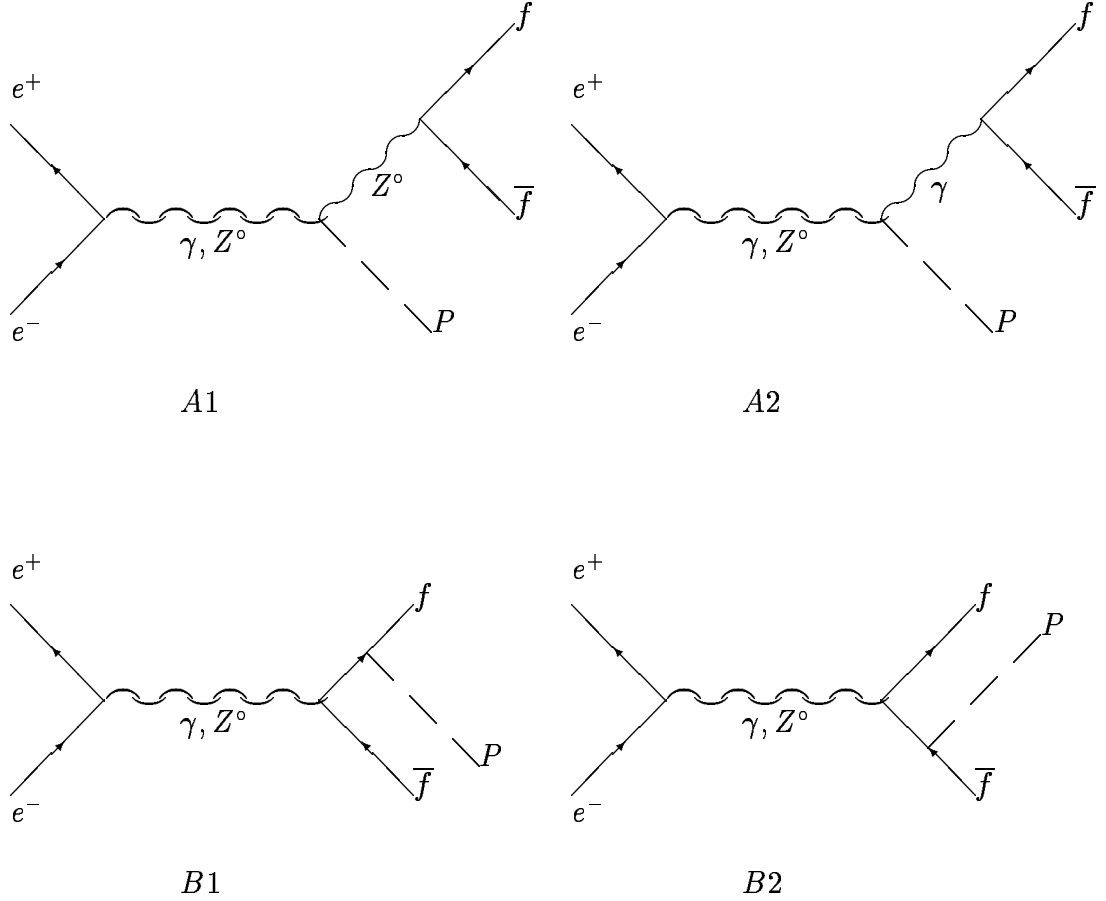}
\vspace{7.0truecm}
\caption[]{\it Feynman diagrams relevant for the processes $e^{+}e^{-}
\ra P\bar ff$ in a model of ETC.}
\protect\label{fig:eePfffd}
\end{figure}

In the processes of eq. (\ref{eePff}), the ETC interactions enter at the
lowest order. The relevant Feynman diagrams are the eight diagrams
represented in fig. \ref{fig:eePfffd}. In the first four diagrams, (A1) and
(A2), the PGB in the final state is produced by the decay of a virtual gauge
boson. Thus, these diagrams are still controlled by the anomalous couplings
between PGBs and standard gauge bosons. In contrast, in the last four
diagrams, (B1) and (B2), the technipion is emitted by one of the two fermions
produced in the final state, and the production of the PGBs occurs via the ETC
interactions.
%___________________________________________________________
%___________________________________________________________

The calculation of the $\sigma (e^{+}e^{-}\ra P\bar ff)$ cross section has
been performed by neglecting the small masses of the electrons in the initial
state.

In the limit in which the finite lifetime of the $Z^0$ gauge boson is also
neglected, the total cross section turns out to be divergent, due to the
contribution of the two diagrams, labeled as (A1) in fig. \ref{fig:eePfffd},
in which the virtual $Z^0$, decaying in the final fermion pair, can be
produced on shell. In order to avoid these divergences, a finite $Z^0$
lifetime has been taken into account.

The same kind of divergences would also appear, in the calculation of the
total cross section, if one neglects the mass of the final fermion pair.
Indeed, in this limit also the virtual photon, entering in the diagrams (A2)
of fig. \ref{fig:eePfffd}, can be produced on shell. In our calculation, the
final fermion masses have been always taken into account. In addition, for
light and massless final leptons (muons and neutrinos), we have imposed a
lower cut on the values of the pair invariant mass, $M_{ff} \geq 1$ GeV.

The results of our calculation show that, at the future lepton collider
experiments, the contributions of the several $e^+e^- \ra P\bar ff$ channels to
the total cross section of PGB production are quite small. For instance, at
the energy of 200 GeV, the values of these cross sections in the LR model are
smaller than $10^{-3}$ pb, corresponding, at LEP II, to less than one event
per year in each channel. In order to observe these processes at LEP
II, a cross section larger by approximately two order of magnitude would be
required, and it is unlikely that this can be achieved in different WTC
models.

\begin{figure}[t]
\ewxy{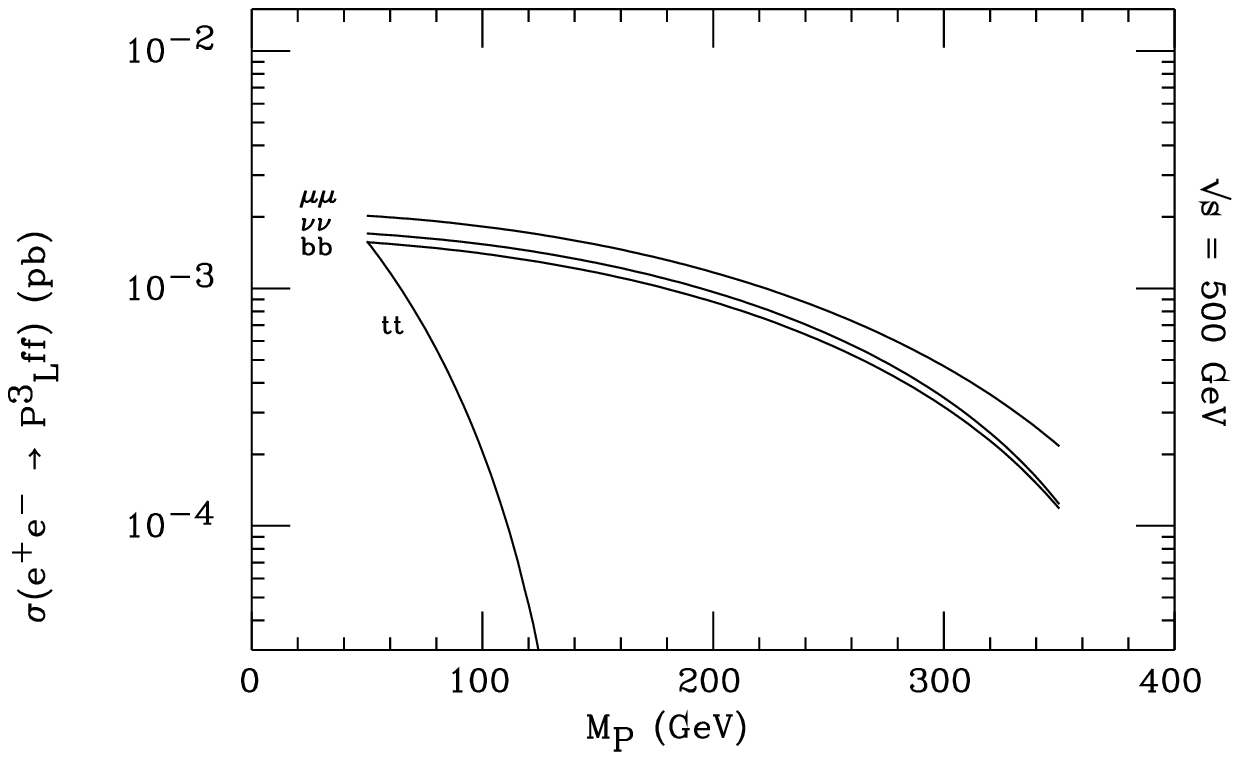}{100mm}
\vspace{3.0truecm}
\caption[]{\it The $\sigma (e^{+}e^{-} \ra P_L^3 \bar ff)$ cross section, in
the LR model, as a function of the PGB mass $M_P$. The energy in the center of
mass is fixed to the value of $500$ GeV.}
\protect\label{fig:eeP3Lff5}
\end{figure}

At the NLC experiment, the $\sigma (e^{+}e^{-}\ra P\bar ff)$ cross sections
are found to be approximately of the same order of magnitude, but the
luminosity of this machine is expected to be much higher than the luminosity
at LEP II. However, it is still unlikely for these processes to be observed.
As an example, we show in fig. \ref{fig:eeP3Lff5} the values of these cross
sections as a function of the PGB mass $M_P$, for $\sqrt{s}=500$ GeV and for
the PGB $P_L^3$ of the LR model. In the figure, we have considered the cases in
which the final fermions are a $\bar tt$, $\bar bb$, $\mu^+ \mu^-$ and $\bar
\nu \nu$ pair. The top quark mass is fixed to the value of 174 GeV.
%___________________________________________________________________________
%___________________________________________________________________________
We see, from fig. \ref{fig:eeP3Lff5}, that all the cross sections, are of the
order, or smaller, than $10^{-3}$ pb, corresponding, at NLC, to a production
rate of only few tenths of events per year. This rate is then found to be
reduced by approximately one order of magnitude for the two PGBs $P^3_Q$ and
$P^0$, whose anomalous TC couplings are of the order of one. We also find that,
via the processes considered in this section, there is no possibility, for
the PGBs of traditional models, to be observed both at LEP II and the NLC
experiments.

Fig. \ref{fig:eeP3Lff5} shows that in all the different final states, by
neglecting the case of final top quarks, the PGBs are produced with
approximately the same probability. This feature reflects the fact that the
PGB production, in these channels, mainly proceeds via the TC anomalous
interactions. Indeed, the contribution of the ETC diagrams, (B1) and (B2) of
fig. \ref{fig:eePfffd}, grows proportionally to the square of the fermion mass.
In these processes, the only case in which the ETC interactions become
relevant is the $e^+e^- \ra P \overline{t}t$ channel. However, as results from
fig. \ref{fig:eeP3Lff5}, at NLC this channel is still strongly suppressed.

\subsection{The production of PGBs in the $e^{+}e^{-}\ra Pe^{+}e^{-}$ channel}

\begin{figure}[t] % produce figure here
\setlength{\epsfxsize}{100mm}\epsfbox[90 390 370 490]{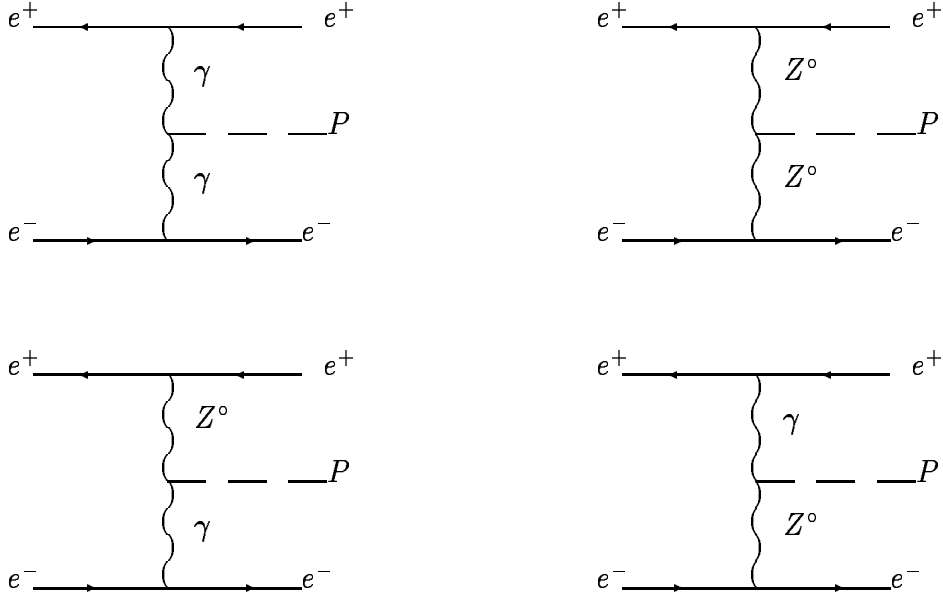}
\vspace{3.5truecm}
\caption[]{\it Feynman diagrams relevant for the $e^{+}e^{-} \ra P e^{+}e^{-}$
channel.}
\protect\label{fig:eePeefd}
\end{figure}

When we consider, among the $e^{+}e^{-}\ra P\bar ff$ processes of the previous
section, the particular case in which an electron-positron pair is produced in
the final state, four new Feynman diagrams enter, at the lowest order, in the
calculation of total cross section. In these diagrams, represented in fig.
\ref{fig:eePeefd}, the PGB is produced by the annihilation of two virtual
gauge bosons, created by the initial electron and positron respectively.
%___________________________________________________________
%___________________________________________________________
These virtual particles can be either photons or $Z^0$ gauge bosons%
\footnote{Similar diagrams, to those reported in fig. \ref{fig:eePeefd}, can
also describe the $e^{+}e^{-}\ra P\overline{\nu }\nu $ channel, that proceeds
through the annihilation of a virtual $W^{\pm }$ pair. However, since the
neutral isoscalar and isovector PGBs are not coupled to $W^{\pm}$ gauge
bosons, we will not discuss this channel.}.

In the limit in which the electron mass is neglected, several sources of
collinear divergences appear in the calculation of the total cross section,
coming from those diagrams of fig. \ref{fig:eePeefd} containing at least one
photon propagator. The degree of these divergences can be linear or
logarithmic.

The effect of the collinear divergences, that, in the actual calculation are
removed by the finite value of the electron mass, is, however, an enhancement
of the resulting cross section. For this reason, we have neglected, in
considering this channel, the contribution, to the total cross section, coming
from the interference between the four diagrams of fig. \ref{fig:eePeefd} and
the eight diagrams considered in the previous section. In addition, we have
found that, because of these collinear divergences, the PGB production in the
$e^+e^-\ra P e^+e^-$ channel, receives the main contribution from the first
diagram in fig. \ref{fig:eePeefd}, contribution that typically accounts for
more than $95\%$ of the total cross section.

In the limit in which only this contribution is taken into account, the total
cross section can be written down in a very compact form. For these two-photon
processes, a convenient procedure to perform the integration on the final
phase space is explained in detail in ref. \cite{BKT}. By indicating with
$p_i$ and $q_i$ ($i=1,2$) the four momenta of the initial and final electrons,
and with $l_i=p_i-q_i$ the momenta of the two virtual photons, one finds:

\begin{equation}
\label{sgggg}
\begin{array}{l}
\sigma _{\gamma \gamma }(e^{+}e^{-}\ra Pe^{+}e^{-})=
\left( \dfrac{A_{P\gamma \gamma }\,N_{TC}\,\alpha ^2}{2\sqrt{2}\pi ^2
F_P\sqrt{n/2}}\right) ^2\ \dint\limits_{-C_L}^{+C_L}d(\cos \theta_1)
\dint\limits_{-C_L}^{+C_L}d(\cos \theta _2)\ \cdot  \\ \qquad \cdot
\dint\limits_0^{2\pi }d\varphi \dint\limits_{M_P}^{E+M_P^2/4E}
\dfrac{d\omega}{\sqrt{\omega ^2-s_0}}\ \dfrac{\vartheta (\omega ^2-s_0)}
{1-\cos \theta }\ \dfrac{\left( E^2-E\omega +s_0/4\right) }{E^2}\
\dfrac B{\left(l_1^2\,l_2^2\right) ^2}
\end{array}
\end{equation}

\noindent
In this formula, $\theta _i$ represents the angle between initial and final
electrons or positrons, $\widehat{p}_i\cdot \widehat{q}_i=\cos \theta _i$,
and $\theta $ is the angle between the two final leptons, $\cos \theta =
\widehat{q}_1\cdot \widehat{q}_2=\sin \theta _1\sin \theta _2\cos \varphi
-\cos \theta _1\cos \theta _2$. The variable $\omega $ represents the energy
of the PGB $P$ and $s_0$ is given by:

\begin{equation}
\label{s0}s_0=\frac{2M_P^2+4\left( E^2-E\omega \right) \left( 1+\cos \theta
\right) }{1-\cos \theta }
\end{equation}

\noindent
The energies of the final leptons, $E_1$ and $E_2$ are expressed, in terms
of $\omega $ and $s_0$, through the relations $E_{1,2}=E-(\omega \pm q)/2$,
where $q=\sqrt{\omega ^2-s_0}$. Finally, the function $B$ comes from the
square of the Feynman amplitude, and has the form:

\begin{equation}
\label{B-fun}B=\frac 14l_1^2\,l_2^2\,B_1-4\,B_2^2+m_e^2\,B_3
\end{equation}

\noindent with:

\begin{equation}
\label{B1B2B3}
\renewcommand{\arraystretch}{1.3}
\begin{array}{cl}
B_1= & (4\,p_1\cdot p_2-2\,p_1\cdot l_2-2\,p_2\cdot l_1+l_1\cdot l_2)^2\ +
 +\ (l_1\cdot l_2)^2\ -\ l_1^2\,l_2^2\ -\ 16\,m_e^4\  \\
B_2= & (p_1\cdot p_2)\,(l_1\cdot l_2)\ -\ (p_1\cdot l_2)\,(p_2\cdot l_1) \\
B_3= & l_1^2\,(2\,p_1\cdot l_2-l_1\cdot l_2)\ +\ l_2^2\,(2\,p_2\cdot
l_1-l_1\cdot l_2)\ +\ 4\,m_e^2\,(l_1\cdot l_2)^2
\end{array}
\renewcommand{\arraystretch}{1.0}
\end{equation}

The total cross section, in eq. (\ref{sgggg}), receives the main contribution
from the region of very small angles $\theta _i$, where, in the limit of
vanishing electron masses, the collinear divergences appear. However, since
this kinematical configuration could be not easily accessible to the
experimental observation, in eq. (\ref{sgggg}) the cut $|\cos \theta _i| \le
C_L$ has been introduced ($C_L=1$ for the total cross section). In our
numerical calculations, we have always considered the case $\theta_i \geq
10^{{\rm o}}$ ($C_L \simeq 0.9848$). This condition reduces the total cross
section by approximately one order of magnitude.

\begin{figure}[t]
\ewxy{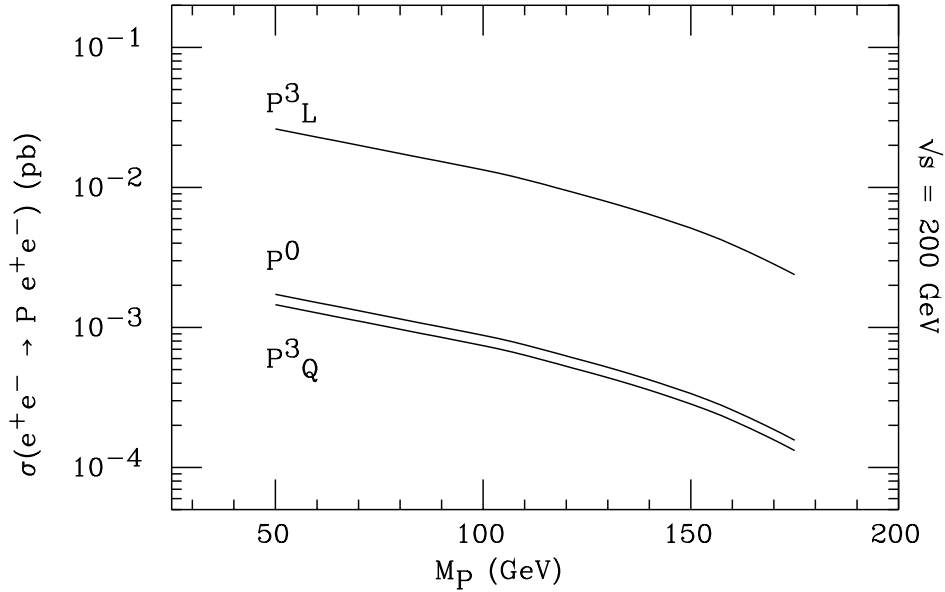}{100mm}
\vspace{3.0truecm}
\caption[]{\it Values of the $\sigma (e^{+}e^{-} \ra P e^{+}e^{-})$ cross
sections, in the LR model, as a function of the PGB mass $M_P$. The energy
in the center of mass is fixed to the value of $200$ GeV.}
\protect\label{fig:eePee2}
\end{figure}

The four dimensional integration in eq. (\ref{sgggg}), or, in general, for
the complete set of diagrams of fig. \ref{fig:eePeefd}, can be performed
numerically. For the particular case of the LR model, the values of the
$\sigma (e^{+}e^{-}\ra Pe^{+}e^{-})$ cross section are shown in fig.
\ref{fig:eePee2} as a function of the PGB mass. The total energy in the center
of mass is fixed, in the figure, at the LEP II value of 200 GeV.
%___________________________________________________________________________
%___________________________________________________________________________
We see, from the figure, that for the PGB $P_L^3$ the values of this cross
section turn out to be of the order of $10^{-2}$ pb, for almost any value of
the PGB mass between 100 and 150 GeV. We also find that, in the cases of the
two PGBs $P^0$ and $P_Q^3$, this rate is reduced by approximately one order of
magnitude.

Thus, in the LR model, the production rate of PGBs, in the $e^{+}e^{-} \ra P
e^{+}e^{-}$ channel, is expected to be of the order of few events per year,
and it is unlikely that these events could be observed at LEP II. Since the
cross section is dominated by the contribution of the first diagram in fig.
\ref{fig:eePeefd}, its value scales essentially like the ratio $(A_{P\gamma
\gamma } N_{TC} / F_P\sqrt{n/2})^2$, as results from eq. (\ref{sgggg}). In
this ratio is contained all the model dependence of this cross section.
Therefore, a larger production can be predicted in the framework of different
WTC models (e.g. models with a smaller TC scale or a larger value of
$N_{TC}$). Since the results of fig. \ref{fig:eePee2} turn out to be probably
at the limit of a possible experimental observation, we then believe that
even this channel is worth an experimental research.

In particular, a clear signal of the $e^{+}e^{-}\ra Pe^{+}e^{-}$ events would
occur if the produced PGB decays into a photon pair. According to our previous
results, this is expected to be the favored decay mode when the ETC coupling
between the PGB and a bottom pair is smaller than one. In this case, the WTC
events would be characterized, in the final state, by an electron pair,
typically emitted at small angles with respect to the direction of the initial
beam, and by a high energetic photon pair. Each of the photons will have
energy larger than $M_P^2/4E$, where $E$ is the energy of the electron beam
(this means $E_{\gamma} \geq 25$ GeV, for $M_P \geq 100$ GeV). In addition,
the photon pair will be monochromatic, with an invariant mass equal to the PGB
mass.

We have estimated that, for these events, the expected Standard Model
background, at LEP II, would be completely negligible, typically smaller by
one order of magnitude with respect to the signal.

In order to compute the signal/background ratio, we have used a Monte Carlo
event generator, kindly provided us by the authors of ref. \cite{Miquel}. With
this code, for any considered value of the PGB mass, we have computed the
number of Standard Model $e^+e^- \ra e^+e^- \gamma\gamma$ events, in which the
photon pair invariant mass turns out to be in the range between $M_P-\Delta M$
and $M_P+\Delta M$, where $\Delta M$ represents the experimental uncertainty
in the measure of the photon invariant mass. The value of $\Delta M$ is
determined by both the uncertainty in the measure of the single photon energy,
that we have assumed to be equal to $\Delta E_{\gamma}/E_{\gamma}=1.2\%$, and
the uncertainty in the measure of the angle between the two photons, that we
have considered to be $\Delta \theta_{\gamma\gamma}=0.5^{\rm o}$. At LEP II,
$\Delta M$ is dominated by the former uncertainty, and we find $\Delta M/M
\simeq 1\%$.

In order to avoid collinear divergences and to select clear events, in
computing the Standard Model background cross section we also have required
final leptons with energy greater than 3 GeV and all the particles well
separated by an angle of at least $10^{\rm o}$. These cuts, applied to the
signal, reduces the corresponding cross sections of approximately $15\%$.
Finally, we have imposed, on the Standard Model background, the same
kinematical constraint of the signal events, namely final leptons with energy
smaller than $E-M_P^2/4E$, and a lepton pair invariant mass smaller than
$2E-M_P$.

After all these cuts have been imposed, we find that the $\sigma(e^+e^- \ra
e^+e^- \gamma\gamma)$ cross section, in the Standard Model, reduces to
approximately $10^{-3}$ pb, corresponding at LEP II to less than 1 event per
year. Thus, in this channel, the TC events are expected to be practically free
of Standard Model background.

\begin{figure}[t]
\ewxy{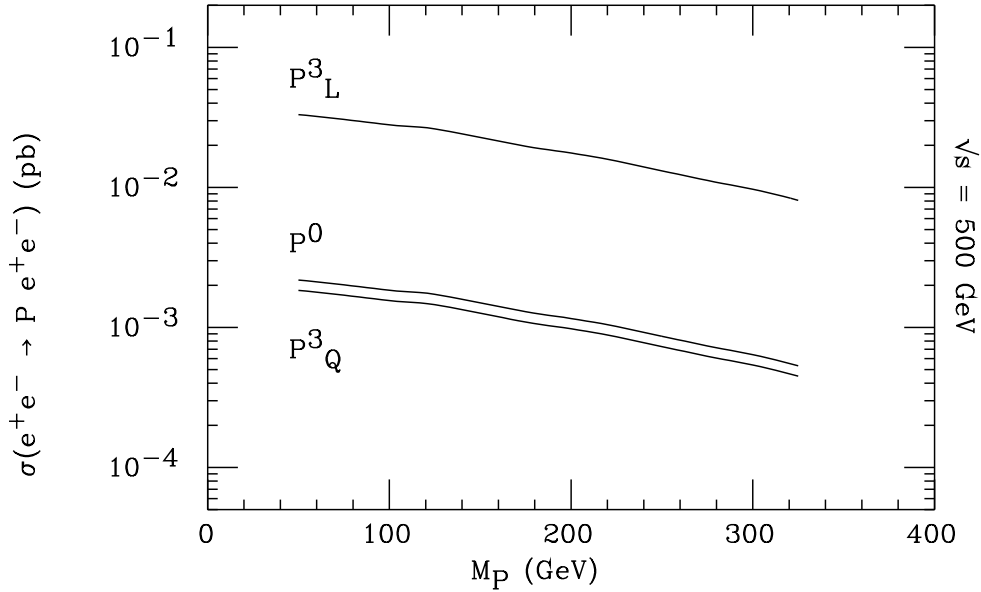}{100mm}
\vspace{3.0truecm}
\caption[]{\it The $\sigma (e^{+}e^{-} \ra P e^{+}e^{-})$ cross sections, in
the LR model, as a function of the PGB mass $M_P$. The energy in the center of
mass is fixed to the value of $500$ GeV.}
\protect\label{fig:eePee5}
\end{figure}

At a 500 GeV NLC experiment, the values of the $\sigma(e^{+}e^{-}\ra Pe^{+}
e^{-})$ cross sections are found to be in the range between $10^{-2}$ and
$10^{-3}$ pb. For the particular case of the LR model, these values are shown
in fig. \ref{fig:eePee5} (they have been computed by always requiring a
minimum angle of $10^{\rm o}$ between initial and final leptons). Thus, we
expect a number of events per year of the order of several hundreds.
%___________________________________________________________________________
%___________________________________________________________________________

Finally, we find that, within the framework of traditional TC models, only few
events per year can be expected, in this channel, at the NLC experiment, by
assuming standard couplings of the order of one as in the case of the FS
model. Moreover, no PGBs of traditional TC models are typically predicted to
be produced at LEP II experiment.

\section{Conclusions}

We now summarize the main results of this paper and present our conclusions.

The existence of a large number of PGBs is a quite general prediction of any
TC/ETC model. In multiscale WTC theories, the lightest of these states have
masses that are typically expected to be larger than 100 GeV. In this paper,
we have studied the production and decay of such particles at the high energy
$e^+e^-$ experiments, LEP II and NLC.

The couplings of neutral PGBs, to ordinary fermions and gauge bosons, can be
written in a form that is, to some extent, model independent. All the model
dependence is explicitly included in the values of the pseudoscalar decay
constants of technipions, in the dimension of the TC gauge group, $N_{TC}$,
and in two classes of model dependent couplings, which, however, are typically
expected to be of order one. Thus, we are able to study quite general
predictions of the theory for the various production and decay rates.

Our results show that, in multiscale WTC theories, because of the existence of
relatively low TC scales, the production of neutral PGBs, in $e^+e^-$
collisions, is significantly enhanced. The corresponding cross sections are
expected to be larger, by one or two orders of magnitude, with respect to the
predictions of traditional single-scale TC models. Thus, despite the typically
large values of PGB masses, these particles could be observed even at the
energy and luminosity of the LEP II experiment.

As an example, we have shown that this production is indeed expected to occur
in the LR model of WTC, provided the PGB masses are not larger than
approximately 150 GeV. In general, a typical condition for this production
to occur at LEP II, with a significant rate, is that the anomalous coupling of
the neutral PGB with a photon pair is larger than one ($A_{P_L^3 \gamma
\gamma} \sim 7$ in the LR model).

At LEP II, the main contributions to the PGB production is expected to come
from the $e^+e^- \ra P \gamma$ channel, and, possibly, with a smaller rate,
from the $e^+e^- \ra P e^+e^-$ channel. For isovector PGBs, we find that the
relevant decay modes are predicted to be into a bottom quark or a photon pair.
In particular, the $P \ra \gamma\gamma$ decay is a typical signature of WTC
dynamics, since it turns out to be usually strongly suppressed in traditional
TC models. At LEP II, in all the channels, by taking into account the
experimental cuts and reconstruction efficiencies, we can expect a number of
events of the order of several tenths per year, assuming an integrated
luminosity of 500 pb$^{-1}$. We have also shown that, in most of the cases,
the distinctive signatures of WTC events allow the Standard Model background
to be reduced to a negligible level. Thus, these particles are worth of an
experimental research.

At the NLC experiment, we expect that the production of neutral PGBs will be
significantly larger, of the order of $10^3$ events per year, assuming a total
energy in the center of mass of 500 GeV and an integrated luminosity of $10^4$
pb$^{-1}$. This production is predicted to occur mainly in the $e^+e^- \ra P
\gamma$, $e^+e^- \ra P Z^0$ and $e^+e^- \ra P e^+e^-$ channels. Instead, no
PGBs are typically expected to be observed, both at LEP II and the NLC
experiment, within the framework of traditional TC models.

\section*{Acknowledgments}

We are very grateful to Ken Lane, for many interesting discussions on the
subject of this paper and for a critical reading of a preliminary version
of the manuscript. We are indebted to Luciano Maiani for having exhorted us to
investigate this topic. Useful discussions with Franco Buccella and Marco
Masetti have been also appreciated. We wish to thank Bing Zhou, for valuable
discussions on the technical features of the LEP II experiment, and Ramon
Miquel, for having provided us with a Monte Carlo generator of the $e^+e^-
\ra e^+e^- \gamma \gamma$ events in the Standard Model. V.L. acknowledges the
support of an INFN post-doctoral fellowship.

\end{document}